\documentclass[12pt,letterpaper]{article}
\usepackage[top=26mm, bottom=40mm, left=26mm, right=26mm]{geometry}
\usepackage{graphicx} 
\usepackage{amsfonts}
\usepackage{amsmath}
\usepackage{xcolor}
\usepackage{tikz}
\usepackage{physics}
\usepackage{hyperref}
\usepackage{datetime}
\usepackage{comment}
\usepackage{ulem}

\numberwithin{equation}{section}

\newdateformat{monthdayyeardate}{%
    \monthname[\THEMONTH]~\THEDAY, \THEYEAR}

\def\g{\gamma}
\def\d{\delta}
\def\G{\Gamma}
\def\O{\Omega}
\def\p{\pi}
\def\ZZ{\mathbb Z}
\def\lp{\left(}
\def\rp{\right)}
\def\lsq{\left[}
\def\rsq{\right]}
\def\Nf{N_{\rm flux}}
\def\nm{n_{\rm massive}}
\def\rmi{{\rm i}}
\def\ll{{\vec l}}
\def\tad{N_{\rm flux}}
\def\ll{{\mathbf{l}}}
\def\nn{{\mathbf{n}}}
\def\mm{{\mathbf{m}}}

\newcommand{\Om}[1]{\chi_{\ll_{#1}}}
\newcommand{\bOm}[1]{\bar{\chi}_{\ll_{#1}}}

\newcommand{\bZ}{\mathbb{Z}}

\newcommand{\bPh}{\mathbf{\Phi}}
\newcommand{\bII}{\beta_{\rm IIB}}
\begin{document}

\begin{titlepage}

\vspace*{-2cm} 

\vspace*{0.8cm} 
\begin{center}
{\Huge Tadpole conjecture  \\
\vspace{0.3cm}
in non-geometric backgrounds}\\

 \vspace*{0.5cm}
Katrin Becker$^{1}$, Nathan Brady$^{1}$, Mariana Gra\~na$^{2}$, Miguel Morros$^{2}$,  \\ [2mm]
  Anindya Sengupta$^{1}$ and Qi You$^{1}$\\

 \vspace*{1.0cm} 
$^1$ {\it Mitchell Institute for Fundamental Physics and Astronomy, Texas A$\&$M University College Station, TX 77843, USA}\\[3mm]

$^2$ {\it Institut de Physique Th\'eorique, Universit\'e Paris Saclay, CEA, CNRS Orme des Merisiers, 91191 Gif-sur-Yvette CEDEX, France}\\[4mm]

{\tt {kbecker@physics.tamu.edu, bradyns@tamu.edu, mariana.grana@ipht.fr, miguel.morros@ens-lyon.fr, anindya.sengupta@tamu.edu, qi\_you@physics.tamu.edu}} \\

\vspace{2cm}
\small{\bf Abstract} \\[3mm]\end{center}

 Calabi-Yau compactifications have typically a large number of complex structure and/or K\"ahler moduli that have to be stabilised in phenomenologically-relevant vacua. The former can in principle be done by fluxes in type IIB solutions. 
 However, the tadpole conjecture proposes that the number of stabilised moduli can at most grow linearly with the tadpole charge of the fluxes required for stabilisation.  We scrutinise this conjecture in the $2^6$ Gepner model: a non-geometric background mirror dual to a rigid Calabi-Yau manifold, in the deep interior of moduli space. By constructing an extensive set of supersymmetric Minkowski flux solutions, we spectacularly confirm the linear growth, while achieving a  higher ratio of stabilised moduli to flux charge 
 than the conjectured upper bound. As a by-product, we obtain for the first time a set of solutions within the tadpole bound where all complex structure moduli are massive. Since the $2^6$ model has no K\"ahler moduli, these show that the massless Minkowski conjecture does not hold beyond supergravity.


\end{titlepage}

\newpage

\tableofcontents

\newpage

\section{Introduction}

Calabi-Yau (CY) flux compactifications have been one of the main focuses of string phenomenology in the past years, for multiple reasons. On one hand, the Standard Model can easily be embedded in the gauge group of compactifications of F-theory or the heterotic theory to four-dimensions (see e.g. \cite{Cvetic:2022fnv,Constantin:2018xkj}), and on the other, they can be used as the building block of de Sitter vacua (see \cite{Cicoli:2023opf} for a review). Furthermore, there are powerful algebraic techniques to build Calabi-Yau manifolds, which have been used to construct thousands of them (see \cite{listCICY} for a list of complete intersection Calabi-Yau three-folds). However, Calabi-Yau manifolds have moduli spaces with typical dimensions of the order of hundreds. Moduli are scalar fields in the four-dimensional effective theory, which, without a potential, would lead to long range ``fifth forces". Since there are very tight experimental bounds on such forces, any phenomenologically-relevant compactification should have no moduli. 

Fluxes are the only known mechanism to generate a potential for the moduli at the perturbative level. However, fluxes have an associated energy-momentum tensor which acts as a source for the Ricci curvature, forcing one to explore other non-Ricci flat backgrounds. There is nonetheless a way to combine NSNS and RR three-form fluxes in type IIB, or more generally four-form fluxes in F-theory, such that Calabi-Yau manifolds (with  a metric that is related to the Ricci flat one by an overall function) are still a solution \cite{Becker:2001pm,Grana:2000jj}. Fluxes then generate a potential for the complex structure moduli and dilaton \cite{Dasgupta:1999ss,Giddings:2001yu}, encoded in the Gukov-Vafa-Witten (GVW) superpotential \cite{Gukov:1999ya}.
Contrarily, K\"ahler moduli can only be stabilised non-perturbatively in this scenario.

Fluxes generate D-brane charges that have to be cancelled globally as required by the tadpole cancellation condition. It is easy to see that in these compactifications the flux generates a positive D3 charge, requiring O-planes for a consistent global solution\footnote{Generically, a supergravity analysis shows that any flux configuration  gives a positive contribution to the four-dimensional energy, making orientifolds a necessary ingredient of four-dimensional Minkowski vacua \cite{Maldacena:2000mw}.}. The total amount of negative D3 charge (coming from O3-planes and/or D7/O7 planes wrapped on 4-cycles) gives an upper bound to the flux charge $\Nf$, leading to a finite number of vacua in a given Calabi-Yau manifold \cite{Denef:2004ze,Grimm:2023lrf}. 

While it was widely believed that turning on a few fluxes with an order one $\Nf$ was enough to fix all moduli, the {\it tadpole conjecture} \cite{Bena:2020xrh}, challenged this common lore by stipulating a linear relation between the number of moduli stabilised (at a generic point in moduli space) and $\Nf$. More concretely, the proposal is that for a large number of moduli 
\begin{equation}
    n_{\rm stab} < \beta\,  \Nf \nonumber \, ,
\end{equation}
with $\beta$ an order 1 number. The refined version of the conjecture proposes an upper bound  $\beta_F <3$ for F-theory compactifications, and  $\bII<3/2$ for type IIB orientifolds.\footnote{Here we are using conventions for orientifolds where the charge $\Nf$ is measured in the covering space, instead of the conventions used in \cite{Bena:2020xrh} where it is measured in the orientifold space. There is a factor of 1/2 in going from the former to the other (see Appendix \ref{app:O} for details), which explains the factor of 1/2 in between $\beta_F$ and $\bII$. We thank Daniel Junghans for bringing this factor to our attention.}. Besides the ``experimental support", the rationale behind this bound comes from F-theory, where the negative O3-charge of 7-branes (typically much larger than that of O3 planes, which was found to be at most 25 in complete intersection Calabi-Yau three-folds \cite{Carta:2020ohw}) is given by the Euler characteristic of the Calabi-Yau 4-fold. For a large number of complex structure moduli, the latter is proportional to 1/4 times the number of  moduli. Then, as long as $\beta_F < 4$ one cannot stabilise all complex structure moduli in F-theory compactifications with a large number of them. 

The tadpole conjecture (including its original bound on $\beta$) was shown to hold in different setups. The original one was F-theory on K3 $\times$ K3 four-folds, as well as in toy models with lower-dimensional cohomology lattices \cite{Bena:2021wyr}. In K3 $\times$ K3, full moduli stabilisation at points in moduli space where the gauge symmetry is purely Abelian, was achieved beyond the tadpole bound, with ${n_{\rm stab}}/{\Nf}=57/25 =2.28$. Other examples of full complex structure moduli stabilisation in IIB at symmetric points in moduli space (points with discrete symmetries) exist, with  $\beta_F$ ranging from $272/124\simeq 2.19$ \cite{Demirtas:2019sip} to $2.86$ \cite{Coudarchet:2022fcl}. Note that these are within the tadpole bound but include a contribution from D7-branes, which either have unstabilised moduli or lead to non-Abelian gauge groups\footnote{On the contrary, full stabilisation at symmetric points (in the deep interior of moduli space) was shown to be in principle possible in supersymmetry-breaking backgrounds with $\Nf$ a priori independent of the number of invariant and non-invariant moduli \cite{Lust:2022mhk}, and an explicit example was given with very low $\Nf$. The stabilisation points are not generic, and as such do not violate the tadpole conjecture, but they lead to discrete, rather than non-Abelian, gauge symmetries. We come back to this point in the conclusions.}. Stabilisation of F-theory on the sextic Calabi-Yau at the Fermat point was studied in detail in \cite{Braun:2020jrx,Braun:2023edp}, where all complex structure moduli are stabilised (beyond the tadpole bound) with the largest $\beta_F=2.94$. Note that in F-theory, stabilisation of complex structure moduli implies stabilisation of both the complex structure moduli of the base manifold and its fiber, as well as the stabilisation of 7-brane moduli. Re. the latter,  the linear relation between the number of 7-brane moduli and $\Nf$ was  proven for any weak Fano base \cite{Bena:2021qty}, where $\beta_F=16/7\simeq 2.29$. Similarly, the linear relation between (any) number of complex structure moduli stabilised at large values or close to conifold point(s) can be very neatly proven using the tools of asymptotic Hodge structure \cite{Grana:2022dfw}. The parameter $\beta$ is generically less than 1, but a more precise bound requires knowledge of the structure of the asymptotic limit. In the examples analysed, a very crude estimate gives $\beta_F$ well below 1.4. 

 By finding a very large set of supersymmetric Minkowski solutions, in this paper we spectacularly confirm the linear growth of $\Nf$ with the number of stabilised moduli, in a completely different setup: non-geometric backgrounds in a substringy and strongly coupled region of moduli space, accessed via their world-sheet description in terms of Landau-Ginzburg (LG) orbifolds \cite{Vafa:1988uu,Vafa:1989xc}. 
The largest value of $\beta$ we find is however more than twice the upper bound $\bII$ proposed in \cite{Bena:2020xrh}, namely $n_{\rm stab}/\Nf=3.4$, calling for a reassessment of this bound in type IIB orientifolds.  

We analyse the so-called $2^6$ Gepner model, corresponding to a tensor  product of 6 minimal models at level 2. Besides being at the deep interior of moduli space, this model is particularly interesting as it has no chiral-antichiral deformations, and consequently the associated cohomology, $h^{1,1}$, is trivial. There are thus no K\"ahler moduli, and as such there is no notion of volume, forbidding a geometric description in this setup. Mirror symmetry relates this model to a (rigid) $T^6/({\mathbb Z}_4 \times {\mathbb Z}_4$) orbifold. Despite being non-geometric, supersymmetric cycles, D-branes and O-planes are well defined \cite{Hori:2000ck}. Furthermore, using perturbative non-renormalisation theorems for the Gukov-Vafa-Witten flux superpotential, as well as the absence of K\"ahler moduli to argue for its non-renormalisation at the non-perturbative level as well, one can claim that this superpotential correctly describes the four-dimensional effective theory also for these non-geometric vacua \cite{Becker:2006ks}. Using then a combination of world-sheet and space-time techniques, \cite{Becker:2006ks} studied complex structure/dilaton moduli stabilisation in the $2^6$ and $1^9$ models, none of which has K\"ahler moduli.      

The absence of K\"ahler moduli led to the belief that all moduli were stabilised in the supersymmetric Minkowski vacua discovered. However, in light of the tadpole conjecture, the $1^9$ model was revisited in \cite{Becker:2022hse, Becker:2023rqi}, showing that full moduli stabilisation actually requires going beyond the tadpole bound. Furthermore, the linear growth proposed by the tadpole conjecture was remarkably verified \cite{Becker:2024ijy}, however with a largest value of $\beta$ that is almost double of the bound proposed in the refined tadpole conjecture, namely $\beta = 114/21 \simeq 5.42$ ($\tad=\tfrac12\, 21$ and $57$ massive moduli). 

 The $2^6$ model (under the specific orientifold we are considering) has the advantage of a smaller ratio $h^{2,1}/|Q_{O3}|=90/40=2.25$ (versus $63/12=5.25$ for the $1^9$), allowing for stabilisation of many more moduli within the tadpole bound. We will  actually find some solutions with {\it all moduli stabilised} in a supersymmetric Minkowski vacua exactly at (as well as slightly below) the tadpole bound. This shows that the massless Minkowski conjecture \cite{Andriot:2022yyj} proposing that in  Minkowski  supergravity  solutions there are always left-over moduli, does not hold beyond the supergravity regime.

The spacetime superpotential $W$ \`a la Gukov-Vafa-Witten, although perturbatively and non-perturbatively exact, is an infinite series in the moduli deformation parameters of the model and one has to resort to an order-by-order computation. The mass matrix, related to the Hessian $\partial_I \partial_J W$, contains information up to order $2$ in this Taylor series expansion. The rank of the Hessian is the number of massive moduli which is a lower bound on the total number of fields stabilised by the superpotential including interaction terms to all orders. The mechanism of stabilising massless fields at higher order in $W$ has been studied in detail for the $1^9$ model most recently in \cite{Becker:2024ijy}. In the paper \cite{higherorder} jointly submitted with ours, the authors study higher-order stabilisation in the $2^6$ model. Remarkably, they find examples where less-than-full-rank solutions get lifted to full stabilisation upon including order $4$ terms in $W$. These are solutions with massless fields but no flat directions in the moduli space, and nicely complement the solutions we present in section \ref{sec:full-rank} where all moduli are stable by acquiring a mass. 

The paper is organised as follows: in section \ref{sec:flux} we review the basic notions of flux compactifications and present the tadpole conjecture. In section \ref{sec:2^6} we present the $2^6$ Landau-Ginzburg orbifold, relegating the detailed computation of its homology and cohomology to Appendix \ref{app:Hodge}. The extensive results obtained using the simplest flux configurations, as well as those using a strategy for the search of solutions with low $\Nf$ are exposed in section \ref{sec:results}. We also present there one solution with all moduli stabilised. Further solutions obtained by an intensive search are given in Appendix \ref{app:2&3ChiSols}. All solutions found are collected in Figure \ref{Fig:data}.

\section{Flux compactifications and tadpole conjecture}
\label{sec:flux}

Even though the $2^6$ Landau-Ginzburg orbifold model does not have a Calabi-Yau analogue, much of the tools of geometric flux compactifications apply, as argued in \cite{Becker:2006ks}. We briefly review those arguments. 

In LG models, orientifold planes and D-branes wrapped on cycles are well understood \cite{Hori:2000ck,Brunner:2003zm,Brunner:2004zd}. The topologically non-trivial 3-cycles, associated to the (co)homology $h^{(2,1)}$ of chiral-chiral marginal deformations of the superpotential, are well defined (see Appendix \ref{app:Hodge} for a brief review). Despite the fact that the world-sheet formulation of RR fluxes is not completely understood, one can resort to a space-time geometric description where each distinct homology cycle can support a quantised flux. Combinations of NS and RR fluxes induce a D-brane charge which must be cancelled globally by D-branes or orientifold planes. In other words, these fluxes must satisfy the same tadpole condition as in geometric compactifications. 

In a geometric setup, odd (even) fluxes generate a potential for the complex structure (K\"ahler) moduli of the Calabi-Yau manifold. In the four-dimensional ${\cal N}=1$ low energy effective theory of Calabi-Yau orientifold compactifications, the potential can be derived from a holomorphic superpotential (one also needs the K\"ahler potential, but in the case we are interested in the K\"ahler potential plays no role). The superpotential can also be computed as the tension of a domain wall separating two vacua with different units of flux, which extends to LG models where D-branes are also well-understood. The superpotential does not receive perturbative corrections, while the non-perturbative instanton corrections depend on the K\"ahler moduli in type IIB compactifications. Since these are absent in the present setup, the classical flux potential is not corrected, and should therefore be applicable to LG models. 

Below we review the basic features of type IIB flux compactifications, pointing out an actual important difference with the geometric setup, which will play no role in the present paper, but is worth emphasizing: the absence of K\"ahler moduli allows for supersymmetric AdS vacua.      

\subsection{Flux compactifications}

We take type IIB backgrounds with NSNS and RR fluxes 
\begin{equation} \label{fluxdef}
    \int_{\Gamma_n} F_3=N_n \in {\mathbb Z}\, ,\quad \int_{\Gamma_n} H_3=M_n \in {\mathbb Z}
\end{equation}
where $\Gamma_n$, $n=1,...,2h^{2,1}+2$, is a basis of integral 3-cycles ($\Gamma_n \in H_3(M,{\mathbb Z})$)  in the covering space $M$ (i.e., before the orientifold projection). Three-form fluxes generate a superpotential, that as argued above is the same as for geometric CY compactifications with O3 planes, given by the Gukov-Vafa-Witten form
\begin{equation}
    \label{eq:stW}
    W=\int G_3 \wedge \Omega \ ,
\end{equation}
where
\begin{equation}
  G_3= F_3 -\tau H_3=(N^n -\tau M^n) \gamma_n  \ ,
\end{equation}
$\Omega$ is the holomorphic $(3,0)$-form and $\gamma_n$ is a  dual basis to the cycles $\Gamma_n$.

The combination of RR and NSNS three-form fluxes has an induced D3-charge $\Nf$, which has to be cancelled by orientifold planes and eventually D-branes:
\begin{equation}
  \Nf 
  = \int_M F_3 \wedge H_3 = \frac{1}{\tau - \bar \tau} \int_M G_3 \wedge\bar G_3= |Q_{O3}| -N_{D3} \ . \label{eq:tadpole1}
\end{equation}
$\Nf$ is then bounded by the absolute value of the (negative) charge of orientifold planes
\begin{equation}
 \, \Nf 
  \le |Q_{O3}| \ . \label{eq:tadpole}
\end{equation}

In this paper we focus on supersymmetric, Minkowski vacua, satisfying $DW=0, W=0$, where the derivative is done with respect to all moduli. For complex structure moduli $t^I$, $I=1,...,h^{2,1}$, this reads
\begin{equation}
   0= D_{t^I} W = \int_M G_3 \wedge \chi_I   \, \quad  
\end{equation}
where $\chi_I$ are the (2,1) forms in cohomology. 

In compactifications on ordinary Calabi-Yau manifolds, the F-term equations for the K\"ahler moduli $D_{\rho} W=0$, require $\langle W\rangle \equiv W_0=0$, and therefore supersymmetric vacua are necessarily Minkowski. There are no K\"ahler moduli in our situation, and the flux superpotential allows for supersymmetric AdS solutions with $W_0 \neq 0$. However, we will be restricting to Minkowski vacua, and the condition $W_0=0$ together with $D_\tau W=0$ sets both $(0,3)$ and $(3,0)$ pieces of the 3-form fluxes to zero\footnote{The condition $D_\tau W$ in the present setup is actually different than in ordinary ISD flux compactifications, since the K\"ahler potential is  $K=-4 \log[ -i (\tau -\bar \tau)]$ instead of the usual one   \cite{Becker:2022hse}. The extra factor of 4 can intuitively be understood as the dilaton piece ``eating" the $-3 \log[ -i (\rho -\bar \rho)]$ of the volume modulus. As a consequence, $D_\tau W$ involves a particular linear combination of (3,0) and (0,3) pieces (to be more precise $(3G+\bar G) \wedge \Omega$), and thus supersymmetric AdS vacua have a (3,0) piece, and as such are not purely ISD.}. In summary,
\begin{equation}
    \label{eq:susyMink}
    G_{\rm Mink} \in H^{(2,1)} \ .
\end{equation}
This flux component is imaginary self-dual and, as such, it is straightforward to show that $\Nf >0$.

\subsection{Moduli stabilisation and tadpole conjecture}

The requirement that $G_3$ should be (2,1) is a complex structure moduli-dependent condition (and also a condition on the axion-dilaton, since it appears in the definition of $G_3$). If, for a given complex structure moduli and dilaton, the flux is (2,1) and there are no infinitesimal deformations that maintain this property, then all moduli are fixed. If on the contrary there are $n$ independent ways of deforming the moduli without altering this property, then the left-over moduli space has dimension $n$. At linear order in the deformations, the number of fixed moduli ($h^{2,1}+1 -n$, where the $+1$ comes from the axion-dilaton) is determined by (half of) the rank of the mass matrix
\begin{equation} \label{massmatrix}
    M= \begin{pmatrix} D_I D_J W & \bar D_{\bar I} D_J W \\ D_{I} \bar D_{\bar J} \bar W & \bar D_{\bar I} \bar D_{\bar J} \bar W \end{pmatrix} \bigg|_{D_IW=0,W=0} = \begin{pmatrix} \partial_I \partial_J W & 0 \\ 0 &  \bar \partial_{\bar I} \bar \partial_{\bar J} \bar W \end{pmatrix} \ .
\end{equation}
The number of massive moduli is then given by
\begin{equation}
    \label{eq:nmassive}
    \nm={\rm rank} \, (\partial_I \partial_J W) \ .
\end{equation}


\vspace{0.2cm}

While it was believed that turning on fluxes for a few cycles should be enough to stabilise all moduli, the {\it tadpole conjecture} \cite{Bena:2020xrh} put forward a direct linear relation between the number of moduli stabilised $n_{\rm stabilised}$ and the flux contribution to the tadpole $\Nf$ for a given flux vacua. More precisely, it says that  at a given $\Nf$, the number of moduli that can be stabilised at a generic point in moduli space is
\begin{equation} \label{tadpoleconj}
   n_{\rm stabilised} < \beta \Nf  \,  \ .
\end{equation}
The refined version says $\bII < 3/2$. As we will show, our results  spectacularly confirm  this linear growth in the $2^6$ LG model, but with a value of $\beta$ that is (slightly above) twice this bound. The  actual bound on $\beta$ has a clear relevance when there is an F-theory uplift, with D7-branes wrapping the 4-cycles associated to K\"ahler moduli. D7-branes provide extra negative contributions to the 3-brane charge, and at the same time have associated moduli. D7-brane moduli and complex structure moduli combine in the F-theory description into the $h^{3,1}$ complex structure moduli of a Calabi-Yau four-fold. Similarly, 3-form fluxes and D7-brane fluxes combine into four-form fluxes, subject to the tadpole cancellation condition
\begin{equation}
    \Nf\le \frac{\chi}{24}=\frac14 (h^{3,1}+ h^{1,1} - h^{2,1} + 8) \sim \frac14 h^{3,1} \, ,
\end{equation}
for $h^{3,1}\gg 1$, and $h^{3,1}\gg h^{1,1},h^{2,1}$. Hence, if $\beta_F <4$ \footnote{See Appendix \ref{app:O} for explanation of the factor of 2 between $\beta_F$ and $\bII$.}, the refined tadpole conjecture $n_{\rm stabilised} < 4 \Nf $, necessarily implies $n_{\rm stabilised}< h^{3,1}$, forbidding full moduli stabilisation in F-theory.

There are two caveats that are worth emphasizing. The first is that moduli can be stabilised at higher orders (i.e., have no quadratic term in the potential, but e.g. a quartic one), 
see \cite{Becker:2022hse} for a discussion in the context of Landau-Ginzburg models, and \cite{Becker:2024ijy, higherorder} for detailed analyses in the $1^9$ and $2^6$ LG models respectively. Considering only massive moduli as we do here, allows to test a weaker form of the conjecture, namely  
\begin{equation}
 \nm \, (\le  n_{\rm stabilised}) < \beta \Nf \ . 
\end{equation}

The second caveat is that we are studying stabilisation at a particular, rather than generic, point in moduli space, namely the Fermat point. 
Therefore, in principle a violation of \eqref{tadpoleconj} in our setup would in principle not provide a counter-example to the tadpole conjecture. We find however that  the linear growth is satisfied, which suggests that points where there is a left-over discrete symmetry, rather than a non-Abelian as suggested in \cite{Bena:2020xrh,Bena:2021wyr}, should still be considered generic.

\section{The $2^6$ model}
\label{sec:2^6}

Orbifolds of Landau-Ginzburg (LG) models are examples of ${\cal N}=(2,2)$ two-dimensional conformal field theories that for central charges $c=\bar c=9$ can realise four-dimensional ${\cal N}=2$ string vacua \cite{Vafa:1989xc}. These models also admit orientifold projections, yielding phenomenologically interesting ${\cal N}=1$ vacua. As in Calabi-Yau compactifications, the chiral-chiral and chiral-antichiral ring of marginal deformations of the SCFT correspond to K\"ahler and complex structure moduli of the four-dimensional theory. Of particular interest are the models that have no (c,c) marginal deformations, as from the Calabi-Yau point of view they would have no K\"ahler moduli. These cannot have a conventional geometric description, as they have $h^{1,1}=0$, and therefore no notion of volume. From the point of view of mirror symmetry, they are mirror to ``rigid" ($h^{2,1}=0$) Calabi-Yau manifolds.      

In more detail, LG orbifold models  are the IR fixed points of  2-dimensional ${\cal N}=(2,2)$ theories of a number $r$ of chiral fields $\Phi_i$ with a relevant deformation coming from a world-sheet superpotential ${\cal W}(\Phi_i)$ that is a homogeneous function (or more generally weighted homogeneous function)\footnote{This means the superpotential satisfies ${\cal W}(\lambda^{\omega_i} \Phi_i)=\lambda^d {\cal W}(\Phi_i)$. Note that ${\cal W}=0$ looks very much like the defining equation for a Calabi-Yau manifold. Indeed, for 5 chiral fields, the corresponding LG model can be thought of as the ``substringy regime" of a Calabi-Yau sigma model, where the CY 3-manifold is defined by the equation ${\cal W}=0$ in ${\mathbb CP}^4$. There are however orbifolds of LG models that do not have a  Calabi-Yau analogue, as the one we analyse here.}. For any given superpotential with this property, there exists a choice of K\"ahler potential ${\cal K}$ such that the theory defined by the action
\begin{equation}
    S=\int d^2 z d^4 \theta {\cal K}(\Phi_i, \bar \Phi_i) + \int d^2z d^2 \theta \, {\cal W}(\Phi_i) + {\rm c.c.}  
\end{equation}
has an IR conformal fixed point \cite{Vafa:1988uu}. For superpotentials of the form 
\begin{equation}
    {\cal W}=\sum_{i=1}^r \Phi_i^{k+2}\, ,
\end{equation}
 the SCFT at the IR fixed point is the Gepner model that is a tensor product of $r$ minimal models at level $k$. These models are referred to as $k^r$ and have a central charge
 \begin{equation} \label{centralcharge}
     c=\sum_{i=1}^r \frac{3k}{k+2} \ .
 \end{equation}
When $c=9$ the 2d SCFTs can, in principle, be used as four-dimensional string vacua. However, one still needs to  demand that the charges of the NS states under the $U(1)$ action that rotates the two supersymmetries is integer. This requires quotienting out the theory by a discrete ${\mathbb Z}_{k+2}$ action. We present below the $2^6$ model that is the focus of this paper. 

\subsection{Orientifold and Hodge numbers}

The $2^6$ Gepner model has six scalar fields and a superpotential given by 
\begin{equation} \label{W2^6}
    \mathcal W = \Phi_1^4 + \ldots + \Phi_6^4~ .
\end{equation}
divided by the $\ZZ_4$ orbifold action 
\begin{equation} \label{orbifold}
    g : \Phi_i \rightarrow \omega \, \Phi_i~, \quad {\rm with} \, \, \omega={\rm e}^{\tfrac{2 \p \rmi}{4}}=\rmi~.
\end{equation}
The IR fixed point corresponds to the tensor product of 6 minimal models at level 2, with total central charge $c=6\times \tfrac32=9$.

From the worldsheet point of view, the orientifold projection is realised by dividing out by worldsheet parity and an involution $\sigma$ such that
\begin{equation}
    {\cal W} (\sigma \Phi_i) = - {\cal W} (\Phi_i)~.
\end{equation}
The simplest such involution, that we choose in the following, is
\begin{equation}
    \label{eq:sigma}
    \sigma\, : \Phi_i \to e^{ \frac{\pi \rmi}{4}} \, \Phi_i~ \, ,
\end{equation}
leading to O3 planes. The total charge of the O3-planes for this involution can be computed using their overlap with the Ramond ground states \cite{Becker:2006ks} and is\footnote{Note this is half of the value reported in \cite{Becker:2006ks}, as we are using different conventions here. See Appendix \ref{app:O} for more details.} 
\begin{equation}
    |Q_{O3}|=20~.
\end{equation}

In the framework of LG/CY correspondence, deformations of the compact space are logically given by appropriate deformations of the worldsheet superpotential. The cohomology ring of the compact space corresponds to the ring of chiral-chiral  primary fields 
\begin{equation}
    {\cal R} = \qty[\mathbb{C} [\Phi_1,...,\Phi_6]/\langle \partial_{\Phi_i} {\cal W} (\Phi_1,...,\Phi_6) \rangle]^{\mathbb{Z}_4}~.
\end{equation}
The marginal deformations are of the form $\Phi_1^{l^1-1}...\Phi_6^{l^6-1}$ with $l^i \in \{1,2,3\}$ and $\sum_i l^i = 2 \mod 4 $. The correspondence between the value of this sum and the degree of the corresponding harmonic forms at the Fermat point (superpotential of the form \ref{W2^6}) is given in Table \ref{tab:complexforms} (see Appendix \ref{app:coho} for details).
\begin{table}[h!] 
    \centering
    \begin{tabular}{|c|c|c|c|c|}
        \hline
        $\sum l_i$ & 6 & 10 & 14 & 18 \\
        \hline
        $H^{(p,q)}$ & $H^{(3,0)}$ & $H^{(2,1)}$ & $H^{(1,2)}$ & $H^{(0,3)}$ \\
        \hline
        Number of forms & $ 1 $ & 90 & 90 & 1 \\
           & $ \Omega $ & $\chi_\ll $ & $ \bar \chi_\ll$ & $\bar \Omega$ \\
        \hline
    \end{tabular}
    \caption{Complex 3-forms at the Fermat point $t^I=0$, where $\ll=(l_1...l_6)$, $l^i \in \{1,2,3\}$}
    \label{tab:complexforms}
\end{table}
For each (2,1) form $\chi_\ll$ there is an associated complex structure deformation $t^\ll$, and the correspondingly deformed superpotential  is
\begin{equation}
    \label{eq:defW}
    {\cal W} (t^\ll) = \sum_{i=1}^6 \Phi_i^4 - \sum_\ll t^\ll \, \bPh^{\ll-1}
\end{equation}
where $\bPh^{\ll-1} \equiv \prod_{i=1}^6 \Phi_i^{l^i-1}$.

It is easy to see that all 3-forms are odd under the orientifold projection \eqref{eq:sigma}. The Hodge numbers for this orientifold model are then 
\begin{equation}
    h^{1,1}=0 ~ , \ \ ~ h^{2,1}_-=h^{2,1}=90~.
\end{equation}
The former implies this Landau-Ginzburg model cannot be associated to an ordinary Calabi-Yau manifold, but rather to the mirror of a rigid one. The mirror manifold is the orbifold $T^6/({\mathbb Z}_4 \times {\mathbb Z}_4)$.

\subsection{Cycles and periods}

As explained in detail in Appendix \ref{app:homo}, a basis of 182 integral three-cycles is given by
\begin{equation} \label{Gammann}
    \Gamma_\nn \in H_3(M,{\mathbb Z}), \ \ \nn = (n_1 \ldots n_6)\, , \ \ n_i = 0, 1, 2 \, , 
\end{equation}
where the $\nn$ correspond to the first 182 integers written in base 3, completing with 0's on the left to build a length six vector. Thus they go from (000000) to (020201). To each $\G_\nn$ we associate its Poincar\'e dual, the real 3-form $\g_\nn$.  The intersection form on the charge lattice can be expressed in matrix form \cite{Becker:2006ks} as
\begin{equation}
\label{eq:Inm}
    \mathbf{I}_{\nn \mm} = \int_{\Gamma_\nn} \gamma_\mm = \Gamma_\nn \cap \Gamma_\mm~.
\end{equation}
See eqn. \eqref{eq:intersection-matrix-form} in Appendix \ref{app:homo} for a more explicit form. The integral of a complex 3-form $\g_\ll$ over any real 3-cycle $\G_\nn$ can be expressed from the worldsheet perspective via the correspondence between Landau-Ginzburg A-branes/chiral primaries $\Phi^{\ll-{\mathbf 1}}$ and spacetime 3-cycles/3-forms $\g_\ll$ (see Appendix \ref{app:homo})
\begin{equation} \label{gammaGamma}
    \int_{\G_\nn} \g_\ll = \int_{\G_\nn} d^6 \Phi \, \, \mathbf{\Phi}^{\ll - \mathbf{1}} e^{-{\cal W}}\overset{t=0}{=}\rmi^{\nn \cdot \ll} ~ \prod_{i} (-1 + \rmi^{l^i})~.
\end{equation}

Let us also mention the existence of the so-called homogeneous basis of (complex) cycles of the compact space $\{C_\ll\}$ that are the support of the (2,1) forms. They have the property 
\begin{equation}
    \int_{C_{\ll'}} \Om{} ~ = ~ 2^{11} ~ \delta_{\ll,\ll'} \prod_{i} (1-\rmi^{l^i})~
\end{equation}
which makes them very convenient for computations. Its explicit construction can be found in Appendix \ref{app:homo}. Besides, these cycles have non-zero intersection only with their complex conjugate, with value
\begin{equation}
    C_{\ll} \cap C_{\ll}^* = 2^{12} \prod_{i} (1-\rmi^{l^i})~.
\end{equation}

\subsection{Fluxes and tadpole cancellation condition}

As argued in section \ref{sec:flux}, RR and NSNS fluxes in LG orientifold models can be treated using usual space-time descriptions. The basis $\{\Gamma_\nn\}$ defined in \eqref{Gammann} supports quantised fluxes as in \eqref{fluxdef}, leading to a complex 3-form flux
\begin{equation}
\label{eq:Gflux-gamma-basis}
    G_3=(N^\nn - \tau M^\nn) \gamma_\nn \ .
\end{equation}
However for convenience we mainly express the fluxes in the complex basis. Since supersymmetric Minkowski solutions require $G_3$ to be (2,1), we should have
\begin{equation}
    \label{eq:Gflux}
    G_3 = \sum_\ll A_\ll ~ \Om{}~.
\end{equation}
The coefficients $A_\ll$ should be such that fluxes are properly quantised: 
\begin{equation}
\label{fluxquant}
    \int_{\G_\nn} G = \sum_\ll \big(A_\ll ~ \rmi^{\nn . \ll} 
    \prod_{i} (1-\rmi^{l_i}) \big) = N_{\nn} - \tau M_{\nn} \ ,  \ \ \ \forall\,  \nn ~. \ \
\end{equation}

Tadpole cancellation condition takes the same form as for geometric compactifications, given in \eqref{eq:tadpole1}. 
For this LG orientifold model we have the upper bound
\begin{equation}
    \Nf\le 40
\end{equation}


Using the Riemann bilinear identity we can write $\Nf$ in terms of the complex coefficients $A_\ll$ in \eqref{eq:Gflux}  
   \begin{align}
     \Nf &= \frac{1}{\tau - \bar{\tau}} \int G_3 \wedge \bar{G}_3 \nonumber \\
     &= \frac{1}{\tau - \bar{\tau}} \sum_{\ll_1, \ll_2, \ll_3} \frac{A_{\ll_1} \bar{A}_{\ll_2}}{C_{\ll_3} \cap C_{\ll_3}^*} \int_{C_{\ll_3}} \Om{1} \int_{C_{\ll_3}^*} \bOm{2}~, \\
   &=  2^{12} \qty(\sum_{\ll = s (111133)} |A_\ll|^2 + \sum_{\ll = s (111223)} 2 ~ |A_\ll|^2 + \sum_{\ll = s (112222)} 4 ~ |A_\ll|^2) \nonumber
\end{align}
where $s$ stands for any permutation of the 6 elements of the vector $\ll$, as these possibilities are all the ones that give $\sum l^i = 10$, and in the last line we have 
substituted $\tau=\rmi$ (we comment on this choice in section \ref{sec:results}).

\subsection{Mass matrix}

As shown in \eqref{eq:nmassive}, for supersymmetric Minkowski solutions the number of massive moduli is given by the rank of the Hessian  of the superpotential. 

Using the flux expression \eqref{eq:Gflux}, the GVW spacetime superpotential \eqref{eq:stW} reads
\begin{equation}
    W = \int G_3 \wedge \O = \sum_{\ll_1, \ll_2} \frac{A_{\ll_1}}{C_{\ll_2} \cap C_{\ll_2}^*} \int_{C_{\ll_2}} \Om{1} \int_{C_{\ll_2}^*} \O~.
\end{equation}
This depends on the dilaton via the definition of $G_3$, such that
\begin{equation}
    \partial_\tau W = - \int H_3 \wedge \O = \frac{1}{\tau - \bar{\tau}} \int (G-\bar{G}) \wedge \O
\end{equation}
The dependence on $t^\ll$ can be found using
\begin{equation}
    \int_{C_{\ll'}} \Om{} = \int_{C_{\ll'}} d^6 \Phi ~ \mathbf{\Phi}^{\ll - {\bf 1}} ~ e^{-{\cal W}(t^\ll)}~.
\end{equation}
where ${\cal W}(t^\ll)$ is the deformed superpotential \eqref{eq:defW}, and  ${\bf 1}=(111111)$. This expression 
is also valid for the holomorphic 3-form $\Omega$, with $ \ll = {\bf 1}$

 We can now compute the second derivative of the superpotential with respect to all moduli. At $\tau=\rmi$ and at the Fermat point $t^\ll=0$ we get
\begin{align}
         \left. \frac{\partial^2}{\partial t^{\ll_1} \partial t^{\ll_2}} W ~  \right|_{t=0} & = \frac{1}{2} \sum_{\ll_3} A_{\ll_3} \delta (\ll_1 + \ll_2 + \ll_3 \! \! \! \! \mod \mathbf{4} ~ , \mathbf{1}) \prod_{i=1}^6 (\rmi^{l_1^i + l_2^i -1} - 1) \G \qty(\frac{l_1^i + l_2^i -1}{4}) \nonumber \\
         \left. \frac{\partial^2}{\partial t^{\ll_1} \partial \tau} W ~  \right|_{t=0} & = \frac{\rmi}{4} \sum_{\ll_2} \qty[A_{\ll_2}^* \delta(\ll_1,\ll_2) - A_{\ll_2} \delta(\ll_1 + \ll_2 \! \! \! \! \mod \mathbf{4} ~ , \mathbf{0})] \prod_{i=1}^6 (\rmi^{l_1^i} - 1) \G \qty(\frac{l_1^i}{4}) \nonumber\\
         \left. \frac{\partial^2}{\partial \tau^2} W ~  \right|_{t=0} & = 0 \ .
\end{align}


\section{Solutions}
\label{sec:results}

In this section, we explicitly present some supersymmetric Minkowski solutions with quantised flux and $\Nf$ below the tadpole bound.

A generic flux $G_3$ can be expressed using the $3$-forms $\gamma_\nn$ that are Poincar\'e dual to the cycles $\Gamma_\nn$. This is the expression \eqref{eq:Gflux-gamma-basis}. We note that this expansion is non-unique since the $\gamma_\nn$ are over-complete to describe $G_3 \in H^{(2,1)}$. However, it is straightforward to choose an independent set of linear combinations and thus parametrise a (2,1) form $G_3$ by the associated \emph{independent} flux numbers. Alternatively, $G_3$ can be expanded in the complex $\Om{}$-basis as in \eqref{eq:Gflux}.  The expansion in the $\Om{}$-basis is rather useful since it manifestly constrains the Hodge-type of the form. These two parametrisations of $G_3$ are obviously equivalent and it is easy to compute the linear transformations that relate the complex coefficients $A_\ll$ to the independent flux numbers, and vice versa.

While generating solutions, we make use of both \eqref{eq:Gflux-gamma-basis} and \eqref{eq:Gflux}. In one approach, we ``turn on" the coefficients $A_\ll$ one by one. That is, we allow a certain number of $A_\ll$'s to be non-zero, while the rest of the $A_\ll$'s are set to zero. The actual value of non-zero $A_\ll$'s is then determined by imposing flux quantisation and the tadpole cancellation bound  $\Nf \leq 40$. We refer to a flux with $n$ non-zero $A_\ll$ coefficients in \eqref{eq:Gflux} as an $n$-$\Om{}$ solution. Proceeding this way, we find a large number of solutions that we present in this section and in Appendix \ref{app:2&3ChiSols}. Already at $n\leq4$, one finds solutions that are on or above the line $\nm = 3 \Nf$ in Figure \ref{Fig:data}. Computational limit notwithstanding, one can chart the entire (finite) set of solutions within the tadpole cancellation bound using this strategy. We did this exhaustively up to $n=3$, and also explored some solutions with $n=4$.   

The other approach is to use \eqref{eq:Gflux-gamma-basis} by first restricting to a $\bZ$-linearly independent spanning set of $\gamma_\nn$'s (or linear combinations thereof). We describe in section \ref{sec:basis-flux-lattice} how to obtain such an integral basis of the supersymmetric flux lattice, following the algorithm designed in \cite{Becker:2024ijy}. The rank of the supersymmetric flux lattice turns out to be\footnote{We will justify this statement shortly.} $182$. Once an integral basis, say $\{\mathbf{e}_i: i = 1, 2, \ldots, 182\}$, is obtained, $G_3$ is \emph{uniquely} expressed as 
\begin{equation}
\label{eq:Gflux-gamma-basis-unique}
    G_3 = \sum_i k^i \, \mathbf{e}_i~,
\end{equation}
where the coefficients $k^i \in \bZ$ are the independent flux numbers we alluded to above. We then generate solutions by ``turning on" a fixed number of the $k^i$, and setting the rest to zero.  We refer to a flux with $n$ non-zero $k^i$ coefficients in \eqref{eq:Gflux-gamma-basis-unique} as a ``level-$n$" solution. In practice, we have generated many solutions by going up to level-$5$, with ``small" values for the non-zero $k^i$'s, typically $|k^i| \leq 2$.

In both of the approaches described above, we restrict ourselves to Minkowski solutions. This is a rank $180$ sublattice in the full lattice (rank $182$) of supersymmetric fluxes. The tadpole cancellation bound $\Nf \leq 40$ picks out a finite region in this infinite sublattice, so one can reach all physical solutions in a finite time. Also, the basis $\{\mathbf{e}_i\}$ we used for generating Minkowski solutions level-by-level can be transformed by judicious $SL(180, \bZ)$ transformations to obtain equivalent integral bases that may be more or less useful for select purposes. We present a couple of bases that we have made use of in this work in the \texttt{GitHub} repository \cite{github}.

Before proceeding to the list of solutions, let us pause to make an important observation. Using e.g. \eqref{gammaGamma} it is not hard to see that quantisation condition can only be satisfied if 
\begin{equation} \label{SL2Ztau}
    \tau=\frac{a \, \omega + b}{c \, \omega +d} \, ,
\end{equation}
where $\omega$ is the phase that defines the orbifold action \eqref{orbifold} (in the $2^6$ model  $\omega=\rmi$), and $a,b,c,d \in {\mathbb Z}$. We make the choice
\begin{equation}
\tau = \omega= \rmi \ .   
\end{equation}



\subsection{$n$-$\Om{}$ solutions}
As described in the introduction to this section, we now set out to generate the $n$-$\Om{}$ solutions.
To exhaustively get all Minkowski solutions with $n$ $\Om{}$'s turned on and $\Nf \leq 40$, one first notes that, at first blush, there are \(\binom{180}{n}\) distinct subspaces of complex dimension $n$ to deal with. However, there is an $S_6$ symmetry of permuting the factors in the tensored model, which induces an action of $S_6$ on this set of subspaces. We use this symmetry to identify the subspaces in the same $S_6$ orbit, and are left with a smaller number of $n$-dimensional subspaces to search through. Note that there can still be some redundancy in the solutions derived from each subspace -- in general there are non-trivial $S_6$ elements that fix a subspace but permute the $\Om{}$'s that span it. The solution set obtained from a subspace can be modded by this residual symmetry to arrive at a list of $S_6$-distinct solutions. We have performed this exercise for $n=1,2,3$ exhaustively, and carried out the $n=4$ case partially. The $1$-$\Om{}$ solutions are presented below, and the complete list of  $2,3$-$\Om{}$ solutions can be found in appendix \ref{app:2&3ChiSols}. In what follows, the rank of the mass matrix of a given solution is denoted by $r$.

\par
\textbf{$1$-$\Om{}$ solutions:}
There are $3\times 4=12$ $S_6$-distinct $1$-$\Om{}$ solutions within the bound $\Nf \leq 40$. These are:
\begin{subequations}
    \begin{align}
        G &= \frac{\rmi^n}{16} ~ \chi_{111223}~, ~ r = 28~, ~ N_{\rm flux} = 32~, \\
        \label{eq:OneChi2}
        G &= \frac{\rmi^n}{32} ~ \chi_{112222}~, ~ r = 36~, ~ N_{\rm flux} = 16~, \\
        G &= \frac{\rmi^n}{32} (1+\rmi) ~ \chi_{112222}~, ~ r = 36~, ~ N_{\rm flux} = 32~,
    \end{align}
\end{subequations}
where $n=0,1,2,3$. The third $1$-$\Om{}$ family of $S_6$ orbits yields solutions whose lowest $\Nf$ value is beyond the tadpole cancellation bound. These are $G = \frac{\rmi^n}{8} \, \chi_{111133}\,$, $r = 21, ~ \Nf = 64$. An overall factor of $\rmi^n$ does not change the physical characteristics of a solution -- the flux remains quantised, and has the same values of $\nm$ (also $n_{\rm stab}$ at all orders in $W$), and $\Nf$. So one should actually not distinguish such fluxes even if they are not related by $S_6$. On the contrary, an overall integer factor would increase $\Nf$, while leaving $\nm$, $n_{\rm stab}$ unaltered, and are not of interest in our chase of solutions with highest $\nm/\Nf$ ratio. These solutions are all well above the bound $\bII<3/2$, the best ratio being ${36}/{16}=2.25$

Already with the help of $1$-$\Om{}$ solutions, we can establish that the rank of the supersymmetric flux lattice, restricted to Minkowski solutions, is $180$ -- the set 
\begin{equation}
    \Bigl\{\frac{\rmi^n}{16} ~ \chi_{s(111223)}, \frac{\rmi^n}{32} ~ \chi_{s(112222)}, \frac{\rmi^n}{8} \, \chi_{s(111133)} \Big| s \in S_6,\, n = 0,1 \Bigr\}
\end{equation}
contains precisely $180$ $\bZ$-linearly independent, integer quantised fluxes. It does not form an integral basis, since its $\bZ$-span is only a proper sublattice. We will present in the next subsection the derivation of an integral basis.

\par
\textbf{$2,3$-$\Om{}$ solutions:}
Continuing in this fashion the use of $S_6$ symmetry to reduce the search space, we derive the complete list of $S_6$-distinct $2,3$-$\Om{}$ solutions with very little effort. These are presented in Appendix \ref{app:2&3ChiSols}, with the code used to generate them made available in \cite{github}.


\subsection{Integral bases for the flux lattice}
\label{sec:basis-flux-lattice}

Ref. \cite{Becker:2023rqi} presented an integral basis of Minkowski fluxes in the $1^9$ LG model, and the algorithm used there was spelled out explicitly in \cite{Becker:2024ijy}. For the readers' convenience, we paraphrase here the argument from \cite{Becker:2024ijy} which we use to obtain an integral basis for Minkowski fluxes in the present model.

We have already described two equivalent parametrisations\footnote{Here, \eqref{eq:Gflux-gamma-basis-unique} is restricted to Minkowski solutions, so there are $180$ independent flux numbers $y^i$ instead of $182$.} of the $3$-form flux:
\begin{equation}
    G_3 = \sum_{\ll: \sum l_i = 10} A_\ll \, \Om{} = \sum_{i=1}^{180} k^i \, \mathbf{e}_i~.
\end{equation}
Let us split the $90$ complex coefficients $A_\ll$ into real and imaginary parts to represent $G_3$ as a vector of $\mathbb R^{180}$. The flux quantisation conditions \eqref{fluxquant}
are constraints on the complex coefficients $A_\ll$. Taking $\tau=\rmi$, and separating \eqref{fluxquant} into real and imaginary parts, we get $182\times 2 = 364$ real constraints on the $180$ real and imaginary parts of $A_\ll$'s. Not all these constraints are linearly independent. In fact, the rank of the system of linear constraints is $180$. A (non-unique) set of independent constraints can be used to express the independent flux numbers $k^i$ in terms of the real and imaginary parts of the complex coefficients $A_\ll$'s. Using the same notation as \cite{Becker:2024ijy},
\begin{equation}
    \sum_{\ll} \mathbf B^i {}_\ll A_\ll = k^i~.
\end{equation}
In the equation above, $A_\ll$ stands (temporarily) for both real and imaginary parts of the complex coefficients $A_\ll$ -- we choose not to introduce new notation for this. The matrix $\mathbf{B}^i {}_\ll$ is real, $180 \times 180$, and has full rank (we already know that the Minkowski flux lattice has rank $180$). Inverting it, we obtain the $A_\ll$ as linear functions of $k^i$: $ A_\ll = A_\ll(k^1, \ldots, k^{180})$. Going back to the flux quantisation equation in the form \eqref{fluxquant}, one can explicitly check that the integers $\{N^\nn, M^\nn \}$ are $\mathbb Z$-linear combinations of the integers $k^i$. 

The columns of $[\mathbf B_{i \ll}]^{-1}$ yield an integral basis of the Minkowski flux lattice. In fact, these correspond to the fluxes $\mathbf{e}_i$ obtained at ``level-$1$" with $k^j = \delta^{ji}$. The explicit form of the basis we constructed in this way can be found in the \texttt{GitHub} repository \cite{github}. Any two integral bases have to be related by a similarity transformation in ${\it SL}(180,\bZ)$.

\begin{figure}[!h]
    \centering
    \includegraphics[scale=1.3]{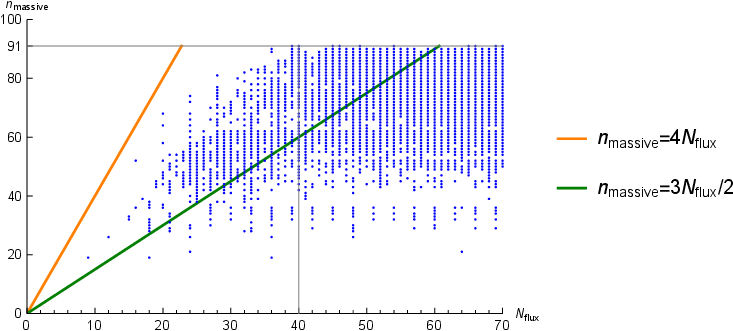}
    \caption{Tadpole and number of massive moduli for solutions found in the $2^6$ model using the strategies described in the text. }
    \label{Fig:data}
\end{figure}

The ``level-$n$" solutions described at the beginning of this section are simply elements of the $\bZ$-span of precisely $n$ of the $\mathbf{e}_i$'s. This way of classifying solutions is highly basis-dependent, but serves us tremendously well in generating data for Figure \ref{Fig:data}. Many of the elements in the integral basis we find this way have large $\Nf$ values, which makes it prone to generating solutions with large $\Nf$. Let us call this a ``long" basis. To find solutions with small $\Nf$ more efficiently, one would try to find a  basis of shorter vectors (where the norm is $\Nf$).  One question is what is the shortest possible vector. The answer  (according to the database of solutions we generated by an intense search) turns out to be 9. These solutions are $9$-$\Om{}$ solutions with $\Nf = 9$ and are elements of the long basis. This is deeply evocative of the shortest vectors in the $1^9$ model \cite{Becker:2023rqi} which are $8$-$\Om{}$ solutions with $\Nf=8$. 
Taking the $\Nf = 9$ vectors in the long basis together with their $S_6$ orbits, we find a   large set of vectors with $\Nf = 9$. The number of linearly independent vectors in this set turns out to be $156$. We then searched for the elements in the original long basis that can be expressed as $\mathbb Z$-linear combinations of these $156$ (possibly) shortest vectors, and replaced them by the latter. The remaining $24$ vectors of the long basis were kept. Let us call this set a ``short" basis.
As a sanity check, we computed the transformation matrix between the long and the short bases described above, and it is indeed in $SL(180, \bZ)$. The short basis is computationally much more efficient in finding solutions with small $\Nf$. We again refer interested readers to the \texttt{GitHub} repository \cite{github} for explicit expressions of these basis vectors.

We have described two strategies of generating solutions -- one organised by the number of $\Om{}$'s turned on, the other organised by the number of independent flux numbers $k^i$'s turned on. The latter strategy has been referred to as ``finding solutions level-by-level". It is facilitated by first building an integral basis of the flux lattice. We have constructed two integral bases -- one long and one short -- which have been used to find numerous data-points at the high\footnote{We filtered out solutions with $\Nf > 70$ from the plot.} and low $\Nf$ regions of Figure \ref{Fig:data} respectively. Figure \ref{Fig:data} also includes the data from the $1,2,3,4$-$\Om{}$ solutions. The highest ratio obtained is $\bII=68/20=3.4$, in the $4$-$\Om{}$ solutions

\begin{align}
    G &= \left \{ \begin{array}{ll}
         \frac{\rmi^m}{64} \lsq (1+\rmi) \chi_{111223} + (1+\rmi) \chi_{113221} + \rmi^n \lp 2 \chi_{221131} \pm \rmi \, \chi_{222112} \rp \rsq  \\[0.2cm]
         \frac{\rmi^m}{64} \lsq (1+\rmi) \chi_{111223} - (1+\rmi) \chi_{113221} + \rmi^n \lp 2 \chi_{221131} \pm \chi_{222112} \rp \rsq ~.
    \end{array}\right. 
\end{align}   




\vspace{1cm}


\subsection{Solutions with all moduli massive}
\label{sec:full-rank}
A very  important question is whether there exists a string vacuum in which all moduli are stabilised by fluxes. This is highly non-trivial, and in fact has never been realised before in a supersymmetric Minkoswki solution, to the best of our knowledge. There is even a \emph{massless Minkowski conjecture} stating that there are always massless moduli in  Minkowski supergravity solutions \cite{Andriot:2022yyj}.

In the present model, we actually find within the tadpole bound many flux configurations that lead to Minkowski vacua with all moduli massive. Below is one such example with $\Nf=40$, and many more can be found in the \texttt{GitHub} repository \cite{github}. 
\begin{equation}
    \begin{aligned}
        G = & \frac{1}{128} \left[ -2 \rmi \chi_{111133}+(1+\rmi) \chi_{111232}- (1+\rmi) \chi_{112231} +2 \rmi \chi_{113131} - (1+\rmi) \chi_{121321} \right. \\
        &  -\chi_{122212} +\rmi \chi_{122221}+(1-\rmi) \chi_{123121} -2 \rmi \chi_{123211}+2 \rmi \chi_{131311} -2 \rmi \chi_{132112} \\
        &  +(1-\rmi) \chi_{132211} +2 \chi_{133111}+(1+\rmi) \chi_{211123}-\chi_{211222}+(1-\rmi) \chi_{211321}+\chi_{212212} \\
        & +2 \rmi \chi_{213211} + \chi_{221212} + \rmi \chi_{222112} + \rmi \chi_{222211} - 2 \chi_{223111} +2 \rmi \chi_{231112} -2 \chi_{232111} \\
        & -2 \chi_{311113} - (1+\rmi) \chi_{311212} +2 \rmi \chi_{311311} -2 \rmi \chi_{312211} -2 \rmi \chi_{321112} +2 \chi_{322111} \left. \right]
    \end{aligned}
\end{equation}
Most of the other full-rank solutions we find also have $\Nf=40$. These do not require mobile D3-branes to satisfy tadpole cancellation. In our dataset, the smallest value of $\Nf$ where full rank is achieved is $\tad=39$. These backgrounds need one D3-brane. Note that these full-rank solutions are far from being the ones with the highest $\nm /\Nf$ ratio, as can be  seen in Figure \ref{Fig:data}.


\section{Conclusions}

We have generated a plethora of supersymmetric Minkowski solutions in an orientifold of the $2^6$ LG model using two separate solution generating strategies described in detail in section \ref{sec:results}. The most significant achievements of this work were two-fold. On one hand, the discovery of quantised flux backgrounds leading to 4d Minkowski spacetime with all moduli massive. On the other, the spectacular confirmation of the linear behaviour of the number of massive moduli relative to the flux charge. 

The possible existence of moduli-free solutions in the non-geometric LG models was anticipated in \cite{Becker:2006ks}, where the authors gave a systematic description of fluxes in these models for the first time. 
However, the check of the actual number of stabilised moduli was only done 20 years later \cite{Becker:2022hse}, motivated by the tadpole conjecture. Surprisingly (or not, depending on the perspective), none of the original solutions turned out to be moduli-free.  

Here we found {\it $13$ distinct $S_6$ orbits of moduli-free Minkowski solutions} with flux charge exactly at the tadpole bound, as well as  $3$ ($S_6$-distinct) solutions that require one extra D3-brane for tadpole cancellation. 

Our findings demonstrate that the massless Minkowski conjecture of \cite{Andriot:2022yyj} does not hold beyond the supergravity regime. To be fair, the original conjecture is stipulated to apply to supergravity solutions, so our solutions are not counter-examples to it. However, they do limit the scope of the conjecture to its original formulation.

Concerning solutions with partial moduli stabilisation, our results  spectacularly confirm the linear behaviour of the tadpole conjecture. As for its refined version, the bound $\bII<3/2$ is severely violated. We found a handful of solutions with a slightly higher ratio $n_{\rm massive}/\Nf=\bII$ than twice the original upper bound proposed in \cite{Bena:2020xrh}. The highest  value found is $\bII=3.4$. It is intriguing that these models violate the bound $\beta_F=2 \bII<4$ suggested from F-theory, and satisfied in all F-theory setups studied so far, as well as in IIB with (world-volume) fluxes stabilising D7 moduli \cite{Bena:2021wyr}. 

There are two subtle points of the tadpole conjecture. One is to pinpoint what counts as a non-generic point in moduli space (such that the conjecture, which excludes them, does not apply). The other is whether it applies to vacua with $W_0 \neq 0$. Regarding the former, points in moduli space that lead to non-Abelian gauge symmetries were excluded from the start, since  stabilisation at such points  can be achieved easily with low $\Nf$\cite{Bena:2020xrh,Bena:2021wyr}\footnote{However, non-Abelian gauge symmetries typically come hand-in hand with their own moduli, which also have to be stabilised. When the symmetries arise from 7-branes, the tadpole conjecture leads to a dead end: on one hand in type IIB it was shown that there is a fixed ratio  between the number total of D7-moduli and the flux charge needed for full stabilisation, namely 16/7$\simeq 2.27$ \cite{Bena:2021qty}, well below the upper bound allowed. On the other, for more general (p,q) 7-branes,  
the most suitable description is that of F-theory, combining brane moduli and bulk complex structure moduli, and full moduli stabilisation requires $\beta\ge4$.} (and the highest the rank of the non-Abelian gauge group, the smallest the $\Nf$ needed to stabilise there \cite{Braun:2023edp}). On the other hand, the question whether points in moduli space with left-over discrete symmetries bypass the tadpole conjecture bound is more subtle. Our results, as well as those in \cite{Becker:2022hse,Becker:2023rqi,Becker:2024ijy,Braun:2023pzd},  show that the conjecture holds at the Fermat point, despite its high level of symmetry. Conversely, Ref. \cite{Lust:2022mhk} finds examples of stabilisation at points with discrete symmetry at the cost of very little $\Nf$, irrespective of the number of moduli. However, these examples have $W_0 \neq 0$, and we suspect this to be the reason for violation of the tadpole conjecture, rather than the presence of discrete symmetries. Indeed, as \cite{Lust:2022mhk} argues, when $W_0\neq0$ the off-diagonal components of the mass matrix \eqref{massmatrix} are non-zero, and furthermore  have full rank. Hence, there can be massless moduli only if there are cancellations between the diagonal and off-diagonal blocks. On the contrary, in the 609 Calabi-Yau manifolds explored in \cite{Lust:2022mhk}, the maximum rank of the mass matrix in $W_0=0$ solutions is less than $0.8$ times the number of non-invariant moduli. These results provide strong evidence that the tadpole conjecture holds despite discrete symmetries, provided $W_0=0$. Note that, in usual compactifications with K\"ahler moduli, perturbative flux solutions with $W_0 \neq 0$ are supersymmetry-breaking (while, once accounting for non-perturbative corrections, these become supersymmetric AdS vacua). In our context, solutions with $W_0 \neq 0$ are supersymmetric AdS  from the start, there are no non-perturbative corrections, and indeed one can easily find examples with all moduli stabilised at low $\Nf$ \cite{Becker:2022hse}. 

A number of obvious extensions to our work can be anticipated. The $\bZ_2 \times S_7$ and $S_6$ symmetries of the $1^9$ and the $2^6$ orientifolds respectively can be exploited in the manner described in this paper, along with clever parallel computation tricks, to classify all physical solutions in these models. This will allow for an exhaustive check of the tadpole conjecture in these models. Short of this, one can also use modern training algorithms to maximise the $\nm / \Nf$ ratio. The relatively small orientifold charge in the $1^9$ model may allow a full classification of all physical solutions with modest computer resources (see \cite{Becker:2024ijy} where the shortest vectors have been fully classified). Some of the present authors are pursuing this problem. Furthermore, Gepner models obtained by tensoring $N$ minimal models have to satisfy $c=9$
in order to serve as a compactification ``manifold" leading to a $4$d spacetime. Visit \href{http://hep.itp.tuwien.ac.at/~kreuzer/CY/}{this link} for a complete list of possibilities satisfying this constraint. The combination of world-sheet and spacetime techniques used in this paper can also be applied to any of these  examples. We leave this exploration for future work.

\section*{Acknowledgements}

We would like to thank Iosif Bena, Antoine Bourget, Andreas Braun, Severin Lust, Ilarion Melnikov, Jakob Moritz, Hector Parra De Freitas, Muthusamy Rajaguru, Johannes Walcher, Alexander Westphal and Timm Wrase for interesting discussions and useful comments. 
We are particularly thankful to Daniel Junghans for pointing out a missing factor of 1/2 in the computation of the flux charge, which has been corrected in this revised version. The work of KB, NB, AS and QY is supported in part
by the NSF grant PHY-2112859. MG is partly supported by the ERC Consolidator Grant 772408-Stringlandscape.

\appendix

\section{Conventions on orientifolds} \label{app:O}

The orientifold action $\Omega \sigma$ involves, besides the world-sheet parity action $\Omega$,  some  ${\mathbb Z}_2$-valued involution $\sigma$, with  $\sigma^2={\rm id}$. 
Under this action, the homology and cohomology bases of the ``covering space" (the space before the quotient by $\sigma$) split into elements that are even or odd. In this paper we are interested in O3 orientifolds. For these, the complex structure moduli of the orientifold space belong to $H^{2,1}_-$ (while the elements of $H^{2,1}_+$ give scalars in vector multiplets), and their dual homology cycles are the support of RR and NSNS 3-form fluxes, which are quantised in the covering space $M$ according to Eq. \eqref{fluxdef}.  In the following Appendix we show how to obtain the homology and cohomology bases for the $2^6$ LG model.

The charge of an Op plane is $-2^{p-5}$ that of a Dp brane. 
The tadpole cancellation condition in the orientifold space is  
\begin{equation}
   \, \Nf^O 
  =\frac12  \int_M F_3 \wedge H_3 = |Q_{O3}| -N_{D3}=\frac14 N_{O3} -N_{D3} \ . \label{eq:tadpole1}
\end{equation}
The factor of $\frac12$ in $\Nf^O$ arises because one should integrate the fluxes in the orientifold space, whose volume is half of the one of the covering space $M$ where quantisation conditions are imposed. $N_{D3}$ counts the number of D3-branes in the orientifold space. 

On the other hand, in the conventions we use in this paper, which are  the ones in \cite{Becker:2006ks}, the whole tadpole cancellation condition is written in the covering space, and each D3-brane is counted together with its image brane as $N^M_{D3}=2$. The charge of an orientifold plane in the conventions of \cite{Becker:2006ks} is $Q^M_{O3}=-\frac12 N^M_{D3}$, and the tadpole cancellation condition reads
\begin{equation}
   \,\Nf\equiv \Nf^M  
  = \int_M F_3 \wedge H_3 = |Q^M_{O3}| -N^M_{D3}=\frac12 N_{O3} -N^M_{D3} =2 \Nf^O\ . \label{eq:tadpole2}
\end{equation}
The orientifold charge of the $2^6$ LG model was computed in \cite{Becker:2006ks} to be $Q^M_{O3}=-40$.

The IIB version of the tadpole conjecture is
\begin{equation}
   n_{\rm stab} < \beta \Nf^O =  \beta \frac12 \Nf \equiv \bII \Nf \ .
\end{equation}
The refined version $\beta <3 $ then implies $\bII < 3/2$.

\section{Homology and cohomology of the $2^6$ model}
\label{app:Hodge}

\subsection{Cohomology}
\label{app:coho}

Harmonic forms are in one-one correspondence with the ground states in the R-sector, with their degree associated to the $U(1)$ R-charges $q, \bar q$. For a CY 3-fold admitting a LG description, states with charge $(q,\bar q)$ are identified  with elements of the Dolbeaut cohomology $H^{\frac32 -q,\tfrac32 + \bar q}$ \cite{Brunner:2004zd}. Below we will count RR ground states and compute the $U(1)$ charges to derive the Hodge diamond. 

The $2^6$ model is a product of 6 minimal models at level $k=2$. A minimal model at level $k$ has $(k+1)$ supersymmetric ground states in the untwisted sector, labelled by a subscript $0$: $|l\rangle_0$, $l = 1, 2, \ldots, k+1$, which correspond to chiral primaries $\Phi^{l-1}$. The RR charges of a state $|l\rangle_0$ are given by
$    q = \bar q = \tfrac l{k+2} - \tfrac 12 $.
We are interested in minimal models at $k=2$, which have the following ground states 

\begin{align}
    |1\rangle_0 &:~~  q =  \bar q = - \tfrac 14  \nonumber \\
    |2\rangle_0 &:~~  q =  \bar q = 0  \\
    |3\rangle_0 &:~~  q =  \bar q =  \tfrac 14   \nonumber
\end{align}

A ${\mathbb Z}_{k+2}$ orbifold action given by ${\rm e}^{-2 \p i J_0}$ results in $k+1$ twisted sectors (labelled by an index $\nu=1,...,k+1$), with a unique ground state in each twisted sector $|0\rangle_{\nu}$ with charge $ q = \bar q = \tfrac \nu{k+2} - \tfrac 12$. For $k=2$ this gives the states

\begin{align}
    |0\rangle_1 &:~~  q = - \bar q = \tfrac 14 \nonumber \\
    |0\rangle_2 &:~~  q = - \bar q = 0  \\
    |0\rangle_3 &:~~  q = - \bar q = - \tfrac 14  \nonumber
\end{align}


We now tensor 6 minimal models at level $k=2$. From the untwisted sector we get the states:
\begin{equation}
\otimes_{i=1}^6 |l^i\rangle_0 \equiv|\ll \, \rangle_0 \ , \quad {\rm with}    q = \bar q = \tfrac 14 \sum_{i=1}^6 l^i - 3~
\end{equation}
Out of these states, we need to project the ones that are invariant under the ${\mathbb Z}_4$ orbifold action, which enforces $ q, \bar q \in \tfrac 12 + \ZZ$. This implies that we only have the possibilities
\begin{equation}
\label{E:HodgeDeg}
    \sum_{i=1}^6 l^i = 6, 10, 14, 18,
\end{equation}
with corresponding charges $q=\bar q=- \tfrac 32, - \tfrac 12, \tfrac 12, \tfrac 32$, Hodge degrees $(3,0), (2,1), (1,2), (0,3)$ respectively. The numbers of states in the classes are $1, 90, 90, 1$ respectively. These exhaust all the harmonic $3$-forms.

From the twisted sectors we have the ground states , $(|0 \rangle_1)^{\otimes6}$, $(|0 \rangle_2)^{\otimes6}$ and $(|0 \rangle_3)^{\otimes6}$ with corresponding charges $q=-\bar q = \tfrac 32, 0, -\tfrac32$. The middle one has integer charge, and does not survive the ${\mathbb Z}_4$ projection. The first one is a $(0,0)$ form, while the last one a $(3,3)$ form. 

In conclusion, the Hodge diamond for the $2^6$ model is given by
$$
\begin{matrix}
  &   &    & 1 &    &   &   \\
  &   & 0  &   & 0  &   &   \\
  & 0 &    & 0 &    & 0 &   \\
1 &   & 90 &    & 90  &   & 1 \\
  & 0 &    & 0 &    & 0 &   \\
  &   & 0  &   & 0  &   &   \\
  &   &    & 1 &    &   &   \\
\end{matrix}
$$

It is not hard to see that all these forms are odd under the orientifold projection \eqref{eq:sigma}, and as such are not projected out in the orientifold model.  

\subsection{Homology}
\label{app:homo}

\begin{figure}[t]
    \centering
    \begin{tikzpicture}
        \draw[->, thick] (0,0) -- (2,0) node[anchor=south]{$\gamma_0$};
        \draw[->, thick] (0,0) -- (0,2) node[anchor=east]{$\gamma_1$};
        \draw[->, thick] (0,0) -- (-2,0) node[anchor=north]{$\gamma_2$};
        \draw[->, thick] (0,0) -- (0,-2) node[anchor=west]{$\gamma_3$};
    \end{tikzpicture}
    \hspace{1cm}
    \begin{tikzpicture}
         \draw[thick] (0.1,0.1) -- (2,0.1) node[anchor=south]{$V_0$};
         \draw[->, thick] (0.1,0.1) -- (0.1,2);
         \draw[thick] (-0.1,0.1) -- (-0.1,2) node[anchor=east]{$V_1$};
         \draw[->, thick] (-0.1,0.1) -- (-2,0.1);
         \draw[thick] (-0.1,-0.1) -- (-2,-0.1) node[anchor=north]{$V_2$};
         \draw[->, thick] (-0.1,-0.1) -- (-0.1,-2);
         \draw[thick] (0.1,-0.1) -- (0.1,-2) node[anchor=west]{$V_3$};
         \draw[->, thick] (0.1,-0.1) -- (2,-0.1);
    \end{tikzpicture}
    \caption{The ``gamma"-contours (left) and V-cycles (right) in the $\Phi$-plane.}
    \label{Fig:gamma}
\end{figure}
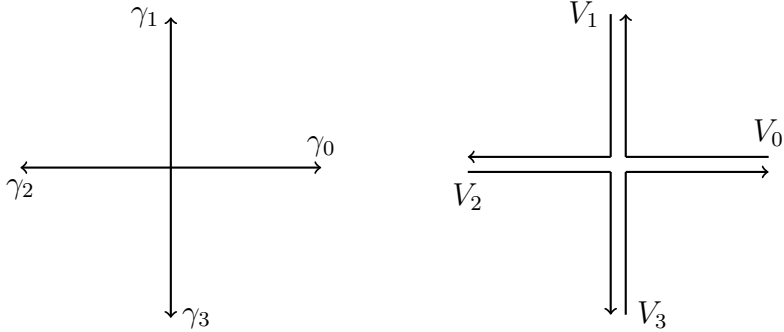

The supersymmetric cycles are the ones wrapped by A and B-branes, corresponding to special Lagrangian (A) and holomorphic (B) cycles. We are interested in the former, which are the cycles that can support 3-form fluxes. 
 A-cycles are given by the preimages of the positive real axis of the worldsheet superpotential. As usual, we start with one minimal model ${\cal W} = \Phi^4$ and then tensor 6 of them. The condition
\begin{equation}
    \textrm{Im} (\mathcal W) = 0~
\end{equation}
gives  four directed contours $\g_0, \g_1, \g_2, \g_3$ whose union makes up this preimage. These are depicted in Figure \ref{Fig:gamma}. A-type D-branes in the minimal model are spanned by the $k+1 =3$ non-compact cycles
\begin{equation}
    V_0 := \g_1 - \g_0, ~~ V_1 := \g_2 - \g_1, ~~ V_2 := \g_3 - \g_2, ~~ V_3 := \g_0 - \g_3~. 
\end{equation}
These cycles clearly satisfy
\begin{equation}
\label{E:Vsum=0}
    V_0 + V_1 + V_2 + V_3 = 0
\end{equation}
and have intersection matrix
\begin{equation}
    V_i \cap V_j = 
    \begin{pmatrix}
        1 & -1 & 0 & 0 \\
        0 & 1 & -1 & 0 \\
        0 & 0 & 1 & -1 \\
        -1 & 0 & 0 & 1
    \end{pmatrix}~.
\end{equation}

For the building block minimal model, the truncated $\ZZ_4$ generator and the intersection matrix in the normalisation used in \cite{Becker:2006ks} are:
\begin{equation}
\label{E:A_I_factor_min_model}
    A = 
    \begin{pmatrix}
        0 & 0 & -1 \\
        1 & 0 & -1 \\
        0 & 1 & -1 
    \end{pmatrix}~,~~~
    I = 
    \begin{pmatrix}
        1 & -1 & 0  \\
        0 & 1 & -1 \\
        0 & 0 & 1 
    \end{pmatrix}~
\end{equation}
The intersection matrix in the full tensored theory is:
\begin{equation}
\label{eq:intersection-matrix-form}
    \mathbf{I} = I^{\otimes 6} (1- A^{\otimes 6} + (A^{\otimes 6})^2 - (A^{\otimes 6})^3)~.
\end{equation}
This matrix is $3^6 \times 3^6$, but has rank $182$. Truncating it to the top-left $182 \times 182$ block yields a full-ranked sub-matrix.

A set of integral three-cycles for the $2^6/\ZZ_4$ model is built by tensoring six $V_n$'s, and then $\ZZ_4$-completing them. Explicitly, these branes are
\begin{align}
\label{E:2^6/Z3-three-cycles}
    \G_{\nn} &= \frac 1{\sqrt 4} \lp V_{\nn} - V_{\nn + {\bf 1}} + V_{\nn + {\bf 2}} - V_{\nn + {\bf 3}} \rp = \frac 1{\sqrt 4} \lp \otimes_i V_{n_i} - \otimes_i V_{n_i + 1} + \otimes_i V_{n_i +2} - \otimes_i V_{n_i +3} \rp~,\\
    & \qquad \qquad \nn = (n_1, \ldots, n_6), \qquad n_i = 0, 1, 2, 3. \nonumber
\end{align}
$\ZZ_4$ acts on $\otimes_i V_{n_i}$ as a tensor product on each of the factors. On a factor $V_n$, it acts as $V_n \to V_{(n+1) ~\textrm{mod}~ 4}$. The set of cycles $\{\G_{\nn}\}$ defined by is linearly dependent. It turns out that one can constrain $n_i$ to $n_i = 0,1,2,$ and further restrict $\nn$'s to be the ternary representations\footnote{written with six ternary digits, padding with zeroes on the left when necessary.} of the first $182$ non-negative integers to obtain an integral basis of three-cycles in the $2^6/\ZZ_4$ orbifold.

There is a nice way to derive the fact that the rank of the lattice $\Lambda$ of three cycles is $182$. The worldsheet superpotential of the $2^6$ model is
\begin{equation}
    \mathcal W = \sum_{i=1}^6 \Phi_i^4 + z^2~,
\end{equation}
where the factor $z$ can be integrated out. However, it is best to not do so because the orbifold action becomes awkward to implement at the level of the $A$-branes. The $\ZZ_4$ action on $\Phi_i, z$ is freely generated by:
\begin{equation}
    \Phi_i \mapsto \rmi \Phi_i~, ~~z \mapsto -z~.
\end{equation}
In the factor $z$-theory, the $A$-branes are two straight wedges, say $V_{z_1}, V_{z_2}$, which coincide but are oppositely oriented:
\begin{equation}
\label{E:Vzsum=0}
    V_{z_1} + V_{z_2} = 0~.
\end{equation}
On a single $\Phi$-factor, the charge lattice $\Lambda_\Phi$ of $A$-branes fits into the exact sequence
\begin{equation}
    0 \to \ZZ \to \ZZ^4 \to  \Lambda_{\Phi} \to 0~,
\end{equation}
where the $\ZZ^4$ is coordinatised by the $V_n$, while the $\ZZ$ can be thought of as the constraint \eqref{E:Vsum=0}, namely $V_0 + V_1 + V_2 + V_3 =0$. On the sixth tensor power of this model, this generalizes to the long exact sequence:
\begin{equation}
    0 \to \ZZ \to 6 \ZZ^4 \to \ldots \to 6 (\ZZ^4)^5 \to (\ZZ^4)^6 \to \Lambda' \to 0~,
\end{equation}
where $\Lambda'$ denotes the lattice of $A$-branes in the tensored model. The rank of $\Lambda'$ before orbifolding is,
\begin{equation}
    {\rm rank} (\Lambda')|_{\rm{pre-orb}} = 4^6 - 6(4^5) + \ldots -6(4) + 1 = (4-1)^6 = 729~.
\end{equation}
$\ZZ_4$ acts on each element of the sequence freely, except for the first $\ZZ$ on which the action is trivial. Hence, in the $\ZZ_4$-orbifold of the tensor product, we ought to divide each summand in the above sum by $4$ except the last one:
\begin{equation}
    {\rm rank} (\Lambda') = \tfrac{4^6 - 6(4^5) + \ldots -6(4)}4 + 1 = \tfrac{(4-1)^6 -1}4 +1 = 183~.
\end{equation}
Similarly, on the $z$-factor, the charge lattice $\Lambda_z$ of $A$-branes fits into the exact sequence
\begin{equation}
    0 \to \ZZ \to \ZZ^2 \to  \Lambda_z \to 0~.
\end{equation}
Tensoring this sequence with the lattice $\Lambda'$ of the $\Phi$-factors, one gets
\begin{equation}
    0 \to \ZZ \to \Lambda' \otimes \ZZ^2 \to  \Lambda \to 0~,
\end{equation}
where $\Lambda$ is the lattice of $A$-branes in the full, orbifolded theory, and the first $\ZZ$ is the relation \eqref{E:Vzsum=0}. And, before orbifolding the $z$-factor
\begin{equation}
    {\rm rank}(\Lambda)|_{\rm{pre-z-orb}} = 183(2) - 1 = 365~.
\end{equation}
The action of $\ZZ_4$ on $\Lambda' \otimes \ZZ^2$ is neither free nor trivial but reduces its dimension by a factor of $2$, and the $\ZZ_4$ action on $\ZZ$ is trivial, hence
\begin{equation}
    {\rm rank}(\Lambda) = \tfrac{183(2)}{2} -1 = 182~.
\end{equation}

The RR charges of the $V_n$'s can be computed as the overlaps $\langle V_n | l\rangle$. Since the Ramond ground states are in one-to-one correspondence to the chiral primary deformations of the superpotential of the minimal model, we have 
\begin{equation}
    |l\rangle \leftrightarrow \Phi^{l-1}~,
\end{equation}
which yields
\begin{equation}
    \langle V_n | l \rangle = \int_{V_n} d \Phi ~ \Phi^{l-1} {\rm e}^{-\Phi^4} = \tfrac{\rmi^l - 1}{4} ~ \G(\tfrac l4) ~ \rmi^{nl}~.
\end{equation}
For the tensored states and cycles, there is a correspondence between Ramond ground states and chiral primaries associated to complex 3-forms $\Om{}$. Consistency with the expression of the intersection matrix $\mathbf{I}$ in \cite{Becker:2006ks} requires the normalization: 
\begin{equation}
     \Om{} ~~ \leftrightarrow ~~ | \ll ~ \rangle ~~ \leftrightarrow ~~ 2^{11} \prod_{i = 1}^6  \tfrac{1}{ \G \lp \tfrac{l^i}4 \rp } ~ \Phi_i^{l^i -1} ~,
\end{equation}
with
\begin{align}
    \langle V_{\nn} | \ll \rangle = \int_{V_{\nn}} \Om{}  &= 2^{11} \prod_{i=1}^6  \int_{V_{n_i}} d\Phi_i ~\frac{\Phi_i^{l^i -1}}{\G \lp \tfrac{l^i}4 \rp} ~ e^{-\Phi_i^4}   = \tfrac 12 ~ \rmi^{\nn \cdot \ll} ~ \prod_{i = 1}^6 (-1 + \rmi^{l^i})~,\\
    {\rm and } ~ \int_{\G_{\nn}} \Om{}   &=  \rmi^{\nn \cdot \ll} ~ \prod_{i = 1}^6 (-1 + \rmi^{l^i}) \label{eq:chiGamma}
\end{align}

We now define another homology basis called the homogeneous basis, which is very adapted to computations involving the basis of complex 3-forms $\Om{}$. We start with a basis of cycles in the minimal model
\begin{subequations}
\begin{align}
    W_1 &:= V_0 + \rmi V_1 - V_2 - \rmi V_3 \\
    W_2 &:= V_0 - V_1 + V_2 - V_3 \\
    W_3 &:= V_0 - \rmi V_1 - V_2 + \rmi V_3~,
\end{align}
\end{subequations}
which can be succinctly written as:
\begin{equation}
    W_l := \sum_{n=0}^3 \rmi^{nl} V_n~, ~~ l = 1,2,3~ \iff V_n = \frac 14 \sum_{l=1}^3 \rmi^{-nl} W_l
\end{equation}
Equation \eqref{E:A_I_factor_min_model} implies the intersection matrix
\begin{equation}
\label{E:Wintersection}
    W_{l'} \cap W_l = 4
    \begin{pmatrix}
        0 & 0 & 1+\rmi  \\
        0 & 1 - \rmi^2 & 0  \\
        1 - \rmi & 0 & 0  
    \end{pmatrix}~ = 4 ~ \delta_{l' + l, 4} ~(1 - \rmi^l)~,
\end{equation}
and therefore
\begin{equation}
\label{E:Cintersection}
    C_{l'} \cap C_l = 4~ \delta_{l' + l, 4}~ (1- \rmi^{-l}) ~ \text{ which simplifies to } ~ C_{l} \cap C_{\bar l} = 4~ (1- \rmi^l)~.
\end{equation}
Reality of the $V_n$ imply $\overline{W_{l}} = W_{4-l} =: W_{\bar l}$, where we have defined the conjugate index $\bar l = 4-l$. In \cite{Becker:2023rqi} some of the present authors had adopted the notation
\begin{equation}
    C_{l} : = W_{4-l} = W_{\bar l}~.
\end{equation}
These cycles satisfy, for $l, l' =1,2,3$, 
\begin{equation}
    \label{E:COmOverlap}
    \int_{W_{l'}} \Phi^{l-1} {\rm e}^{-\Phi^4} = \delta_{l' + l, 4} ~ (-1 + \rmi^l) \G(\tfrac l4) ~ \Leftrightarrow ~ \int_{C_{l'}} \Phi^{l-1} {\rm e}^{-\Phi^4} = \d_{l, l'} ~ (-1 + \rmi^l) \G(\tfrac l4)~.
\end{equation}

Just like before, to obtain the homogeneous cycles of the full theory we tensor six cycles from the minimal models such that
\begin{equation}
    C_{\ll} ~ = ~ \bigotimes_{i=1}^6 C_{l^i} ~ \text{ and } ~ C_{\ll}^* ~ = ~ \bigotimes_{i=1}^6 C_{\bar{l}^i} ~,
\end{equation}
with symplectic intersection matrix
\begin{equation}
    C_{\ll} \cap C_{\ll}^* = \prod_{i=1}^6 C_{l} \cap C_{\bar l} = 2^{12} \prod_{i=1}^6 (1-\rmi^{l^i})~.
\end{equation}
The integration of the complex 3-forms $\Om{}$ over the homogeneous cycles now has the convenient form
\begin{equation}
    \int_{C_{\ll'}} \Om{} ~ = ~ 2^{11} ~ \delta_{\ll,\ll'} \prod_{i=1}^6 (1-\rmi^{l^i})~.
\end{equation}

\section{Two and three $\chi$ solutions}
\label{app:2&3ChiSols}

We present in this appendix a complete list of $S_6$-distinct $2,3$-$\Om{}$ solutions. For $2$-$\Om{}$ solutions, we have modded out the solution set in $\textrm{span}_{\bZ}\{\Om{\ll_1}, \Om{\ll_2}\}$ by the stabilizer subgroup (under the action of $S_6$) of the set $\{\Om{\ll_1}, \Om{\ll_2}\}$, and only enumerate representatives from distinct $S_6$ orbits. This complicates the presentation of the solutions since some patterns are lost after modding by the residual $S_6$. To avoid the proliferation of equations, we do not perform this final modding out for the $3$-$\Om{}$ solutions, and present an $S_6$-redundant set that can be expressed more compactly.

\vspace{0,4cm}
\textbf{$2$-$\Om{}$ solutions:}
The list of $2$-$\Om{}$ solutions below has been broken into $24$ families. In general, $\Nf$ is a homogeneous quadratic in the variables $k^i$, $i=1,2,\ldots, 180$, but for $2$-$\Om{}$ solutions it reduces to a quadratic in $4$ integer variables. The $24$ families below are categorised by the form of this quadratic in $4$ variables.
\begin{subequations}
    \begin{align}
        G &= \left\{ \begin{array}{ll}
        \frac{\rmi^m}{16} \lp \chi_{111133} + \chi_\ll \rp \\[0.2cm]
        \frac 1{16} \lp - \chi_{111133} + \chi_\ll \rp \\[0.2cm]
        \frac {\rmi}{16} \lp - \chi_{111133} + \chi_\ll \rp
        \end{array} \right. 
        ~, ~~ \ll = \left\{
                \begin{array}{ll}
                  111313,~~ r = 28 \\
                  113311,~~ r = 36
                \end{array}
                \right., ~~~ N_{\rm flux} = 32  \\
        G &= \rmi^m \lp \tfrac{(1+\rmi)}{16} \chi_{111133} + \rmi^n \tfrac{(1+\rmi)}{64} \chi_{\ll}  \rp
        ~, ~~ \ll = \left\{
                \begin{array}{ll}
                  112222,~~ r = 44\\
                  122212,~~ r = 50
                \end{array}
                \right., ~~~ N_{\rm flux} = 40  \\
        G &= \rmi^m \lp \tfrac{(1+\rmi)}{16} \chi_{111133} + \rmi^n \tfrac{(1+\rmi)}{64} \chi_{222211}  \rp
        ~, ~~ r = 55,   ~~~ N_{\rm flux} = 40 \\
        G &= \left\{ \begin{array}{ll}
        \frac{1}{32} \lp (-1-\rmi) \chi_{111223} + \rmi^m (1 - \rmi) \chi_\ll \rp \\[0.2cm]
        \frac{1}{32} \lp (-1+\rmi) \chi_{111223} + \rmi^q (1-\rmi) \chi_\ll \rp \\[0.2cm]
        \frac{1}{32} \lp (1-\rmi) \chi_{111223} + \rmi^s (1 - \rmi) \chi_\ll \rp \\[0.2cm]
        \frac{1}{32} (1+\rmi) \lp \chi_{111223} +  \chi_\ll \rp
        \end{array}
        \right., ~ \ll = \left\{
                \begin{array}{ll}
                  111232,~~ r = 28\\
                  112132,~~ r = 38
                \end{array}
                \right., ~~~ N_{\rm flux} = 32 \\
        G &= \left\{ \begin{array}{ll}
        \frac{1}{32} \lp (-1-\rmi) \chi_{111223} + \rmi^m (1 - \rmi) \chi_\ll \rp \\[0.2cm]
        \frac{1}{32} \lp (-1+\rmi) \chi_{111223} + \rmi^q (1-\rmi) \chi_\ll \rp \\[0.2cm]
        \frac{1}{32} \lp (1-\rmi) \chi_{111223} + \rmi^s (1 - \rmi) \chi_\ll \rp \\[0.2cm]
        \frac{1}{32} (1+\rmi) \lp \chi_{111223} +  \chi_\ll \rp
        \end{array}
        \right., ~ \ll = \left\{
                \begin{array}{ll}
                  112123,~~ r = 28\\
                  122113,~~ r = 38
                \end{array}
                \right., ~~~ N_{\rm flux} = 32 \\
        G &= \rmi^m \lp \tfrac{(1+\rmi)}{32} \chi_{111223} + \rmi^n \tfrac{(1+\rmi)}{64} \chi_\ll  \rp
        ~, ~~ \ll = \left\{
                \begin{array}{ll}
                  112222,~~ r = 38\\
                  122122,~~ r = 46\\
                  222112,~~ r = 56
                \end{array}
                \right., ~~~ N_{\rm flux} = 24  \\
        G &= \rmi^m \lp \tfrac{(1+\rmi)}{32} \chi_{111223} + \rmi^n \tfrac{(1+\rmi)}{32} \chi_\ll  \rp
        ~, ~~ \ll = \left\{
                \begin{array}{ll}
                  112231,~~ r = 38\\
                  122131,~~ r = 46
                \end{array}
                \right., ~~~ N_{\rm flux} = 32 \displaybreak \\
        G &= \left\{ \begin{array}{ll}
             \frac 1{32} \lp (-1 - \rmi) \chi_{111223} + \rmi^m (1-\rmi) \chi_\ll \rp  \\[0.2cm]
             \frac 1{32} \lp (-1 + \rmi) \chi_{111223} + \rmi^q (1-\rmi) \chi_\ll \rp \\[0.2cm]
             \frac 1{32} \lp (1 - \rmi) \chi_{111223} + \rmi^s (1-\rmi) \chi_\ll \rp \\[0.2cm]
             \frac 1{32} \lp (1 + \rmi) \chi_{111223} + (1 + \rmi) \chi_\ll \rp
        \end{array}
        \right., ~ \ll = \left\{
                \begin{array}{ll}
                  123121,~~ r = 44\\
                  223111,~~ r = 50
                \end{array}
                \right., ~~~ N_{\rm flux} = 32 \\
        G &= \left\{ \begin{array}{ll}
             \frac {\rmi^m}{32} \lp \chi_{111223} + \chi_{113221} \rp  \\[0.2cm]
             \frac {\rmi^s}{32} \lp -\chi_{111223} + \chi_{113221} \rp 
        \end{array}
        \right.~, ~~ r = 36 ~~~ N_{\rm flux} = 16  \\
        G &= \left\{ \begin{array}{ll}
        \frac{1}{32} \lp (-1-\rmi) \chi_{111223} + \rmi^m (1 - \rmi) \chi_{113221} \rp \\[0.2cm]
        \frac{1}{32} \lp (-1+\rmi) \chi_{111223} + \rmi^q (1-\rmi) \chi_{113221} \rp \\[0.2cm]
        \frac{1}{32} \lp (1-\rmi) \chi_{111223} + \rmi^s (1 - \rmi) \chi_{113221} \rp \\[0.2cm]
        \frac{1}{32} (1+\rmi) \lp \chi_{111223} +  \chi_{113221} \rp
        \end{array}
        \right.~, ~~ r = 36 ~~~ N_{\rm flux} = 32 \\
         G &=  \rmi^m \lp \tfrac{(1+\rmi)}{32} \chi_{111223} + \rmi^n \tfrac{(1+\rmi)}{64} \chi_\ll  \rp
        ~, ~~ \ll = \left\{
                \begin{array}{ll}
                  122221,~~ r = 44\\
                  222121,~~ r = 54
                \end{array}
                \right., ~~~ N_{\rm flux} = 24  \\
        G &= \left\{ \begin{array}{ll}
        \frac{1}{64} \lp (-1-\rmi) \chi_{112222} + \rmi^m (1 - \rmi) \chi_\ll \rp \\[0.2cm]
        \frac{1}{64} \lp (-1+\rmi) \chi_{112222} + \rmi^q (1-\rmi) \chi_\ll \rp \\[0.2cm]
        \frac{1}{64} \lp (1-\rmi) \chi_{112222} + \rmi^s (1 - \rmi) \chi_\ll \rp \\[0.2cm]
        \frac{1}{64} (1+\rmi) \lp \chi_{112222} +  \chi_\ll \rp
        \end{array}
        \right.~, ~~ \ll = \left\{
                \begin{array}{ll}
                  121222,~~ r = 36\\
                  \textcolor{blue}{221122,~~ r = 52}
                \end{array}
                \right., ~~~ N_{\rm flux} = 16  \\
        G &= \left\{ \begin{array}{ll}
        \frac{1}{32} \lp \rmi^m ~.~ \rmi~ \chi_{112222} +  \chi_\ll \rp \\[0.2cm]
        \frac{1}{32} \lp \rmi^q ~.~ \rmi~ \chi_{112222} + \rmi~ \chi_\ll \rp \\[0.2cm]
        \frac{1}{32} \lp \rmi^s ~.~ (-1)~ \chi_{112222} - \rmi ~ \chi_\ll \rp \\[0.2cm]
        - \frac{1}{32}  \lp \chi_{112222} +  \chi_\ll \rp
        \end{array}
        \right.~, ~~ \ll = \left\{
                \begin{array}{ll}
                  121222,~~ r = 36\\
                  221122,~~ r = 52
                \end{array}
                \right., ~~~ N_{\rm flux} = 32
    \end{align}
\end{subequations}
where $m,n = 0,1,2,3$, $q = 0,1,2$, and $s=0,1$.

\vspace{0,3cm}
 The solution in blue has the highest ratio $\bII=3.25$, more than twice above the upper bound of the refined tadpole conjecture.

\newpage
\vspace{0,2cm}
\textbf{$3$-$\Om{}$ solutions:}
\vspace{0,3cm}
The $3$-$\Om{}$ solutions below are organised by their $\Nf$ values.

\textbf{$\tad = 12$:}
\begin{equation}
\begin{aligned}
    G &= \frac{\rmi^m}{64} \lsq (1+\rmi) \chi_{\ll_1} + \rmi^n \lp  \chi_{\ll_2} + \rmi^n (1-\rmi) \chi_{\ll_3}  \rp \rsq  \\
    (\ll_1, \ll_2, \ll_3) &= \left\{
    \begin{array}{ll}
        (111223, 112222, 113221), 
    \end{array}~\right. 
\end{aligned}
\end{equation}
with rank $26$.


\textbf{$\tad = 20$:}
\begin{equation}
\begin{aligned}
    G &= \frac{\rmi^m}{32} \lsq (1+\rmi) \chi_{\ll_1} + \rmi^n \lp \tfrac 12 \chi_{\ll_2} + \rmi^n (1-\rmi) \chi_{\ll_3} \rp \rsq \\
    (\ll_1, \ll_2, \ll_3) &= \left\{
    \begin{array}{ll}
    (111133, 112222, 113311),
    \end{array}\right.
\end{aligned}
\end{equation}
with rank $48$.


\textbf{$\tad = 24$:}
\begin{equation}
\begin{aligned}
    G &= \frac{\rmi^m}{32} \lsq (1+\rmi) \chi_{\ll_1} + \rmi^n \lp  \chi_{\ll_2} + \rmi^n (1-\rmi) \chi_{\ll_3}  \rp \rsq \\
    (\ll_1, \ll_2, \ll_3) &= \left\{
    \begin{array}{ll}
        (111133, 111223, 111313), ~~ (111133, 113221, 111313),
    \end{array}~\right.
\end{aligned}
\end{equation}
with ranks $21$, $44$.

\begin{equation}
\begin{aligned}
    G &= \frac{\rmi^m}{64} \lsq 2 \chi_{\ll_1} + \rmi^n (1+\rmi) \chi_{\ll_2} \pm 2 \rmi  \chi_{\ll_3} \rsq \\
    (\ll_1, \ll_2, \ll_3) &= \left\{
    \begin{array}{ll}
        (111223, 112222, 131221), ~~ (111223, 122122, 113221), \\
        (111223, 221122, 113221), ~~ (111223, 222112, 113221), \\
        (111223, 221221, 113221),
    \end{array}~\right.
\end{aligned}
\end{equation}
with ranks $44$, $44$, $58$, $62$, $52$.

\begin{equation}
    G = \left\{
    \begin{array}{ll}
         \frac{\rmi^m}{64} \lsq 2 \chi_{111223} + \rmi^n \lp (1+\rmi)  \chi_{112222} + 2 \rmi^{n+1}  \chi_{113221} \rp \rsq~, ~~~ r = 26,  \\[0.2cm]
         \frac{\rmi^m}{64} \lsq 2 \chi_{111223} + \rmi^n \lp (1+\rmi)  \chi_{112222} - 2 \rmi^{n+1}  \chi_{113221} \rp \rsq~, ~~~ r = 38~. 
    \end{array}\right.
\end{equation}


\textbf{$\tad = 28$:}

\begin{equation}
\begin{aligned}
    G &= \frac{\rmi^m}{64}\left\{ \begin{array}{ll}
         \lsq (1+\rmi) \chi_{\ll_1} + \rmi^n \lp \chi_{\ll_2} + \rmi^n (-3-\rmi) \chi_{\ll_3} \rp \rsq  \\
         \lsq (3+\rmi) \chi_{\ll_1} + \rmi^n \lp \chi_{\ll_2} + \rmi^n (-1-\rmi) \chi_{\ll_3} \rp \rsq \\
         \lsq (1+3\rmi) \chi_{\ll_1} + \rmi^n \lp \chi_{\ll_2} + \rmi^n (1+\rmi) \chi_{\ll_3} \rp \rsq \\
         \lsq (1+\rmi) \chi_{\ll_1} + \rmi^n \lp \chi_{\ll_2} + \rmi^n (1+3\rmi) \chi_{\ll_3} \rp \rsq \\
         \lsq (1+\rmi) \chi_{\ll_1} + \rmi^n \lp (1+2\rmi) \chi_{\ll_2} + \rmi^n (1-\rmi) \chi_{\ll_3} \rp \rsq \\
         \lsq (1+\rmi) \chi_{\ll_1} + \rmi^n \lp (2+\rmi) \chi_{\ll_2} + \rmi^n (-1+\rmi) \chi_{\ll_3} \rp \rsq
    \end{array}\right. \\
    (\ll_1, \ll_2, \ll_3) &= \left\{
    \begin{array}{ll}
        (111223, 112222, 113221), ~~
    \end{array}~\right.
\end{aligned}
\end{equation}
with rank $38$.


\textbf{$\tad = 32$:}
\begin{equation}
\begin{aligned}
    G &= \frac{\rmi^m}{32} \lsq (1+\rmi) \chi_{\ll_1} + \rmi^n \lp \chi_{\ll_2} \pm \rmi \chi_{\ll_3} \rp \rsq \\
    (\ll_1, \ll_2, \ll_3) &= \left\{
    \begin{array}{ll}
    (111223, 111232, 113212), ~~ (111223, 112132, 112312), \\
    (111223, 112123, 112321), ~~ (112123, 111223, 131221), \\
    (122113, 111223, 113221), ~~ (221113, 111223, 113221), \\
    (112132, 111223, 131221), ~~ (132112, 111223, 113221), \\
    (231112, 111223, 113221), ~~ (112231, 131221, 111223), \\
    (221131, 113221, 111223), ~~ (111223, 112231, 132211), \\
    \{(113221, 131221, 111223), (131221, 111223, 113221), (111223, 113221, 131221)\}, \\
    (231121, 113221, 111223), ~~
    \end{array}\right.
\end{aligned}
\end{equation}
with ranks $38$, $44$, $38$, $46$, $48$, $58$, $50$, $52$, $60$, $46$, $56$, $50$, $\{44\}$, $54$. 

\begin{equation}
\begin{aligned}
    G &= \frac{\rmi^m}{64} \lsq 2 \chi_{\ll_1} + \rmi^n (1+\rmi) \chi_{\ll_2} + \rmi^p (1+\rmi) \chi_{\ll_3} \rsq \\
(\ll_1, \ll_2, \ll_3) &= \left\{
    \begin{array}{ll}
    \{(112222, 121222, 122122), (121222, 122122, 112222), (122122, 112222, 121222) \}, \\
    \{(112222, 121222, 212122), (121222, 212122, 112222), (212122, 112222, 121222) \}, \\
    \{(112222, 121222, 222112), (121222, 222112, 112222), (222112, 112222, 121222) \}, \\
    \{(112222, 221122, 222211), (221122, 222211, 112222), (222211, 112222, 221122) \}, \\
    \{(112222, 121222, 211222), (121222, 211222, 112222), (211222, 112222, 121222) \}~,
    \end{array}\right.
\end{aligned}
\end{equation}
with ranks $\{44\}$, $\{52\}$, $\{58\}$, $\{66\}$, $\{42\}$.

\begin{equation}
\begin{aligned}
    G &= \frac{\rmi^m}{32} \lsq \chi_{\ll_1} + \rmi^n \chi_{\ll_2} \pm \chi_{\ll_3} \rsq \\
    (\ll_1, \ll_2, \ll_3) &= \left\{
    \begin{array}{ll}
        (111223, 112222, 131221), ~~ (111223, 122122, 113221), \\
        (111223, 221122, 113221), ~~ (111223, 222112, 113221), \\
        (111223, 221221, 113221),
    \end{array}~\right.
\end{aligned}
\end{equation}
with ranks $44$, $44$, $58$, $62$, $52$.

\begin{equation}
    G = \left\{
    \begin{array}{ll}
         \frac{\rmi^m}{32} \lsq \chi_{111223} + \rmi^n \lp \chi_{112222} + \rmi^n  \chi_{113221} \rp \rsq~, ~~~ r = 36,  \\[0.2cm]
         \frac{\rmi^m}{32} \lsq \chi_{111223} + \rmi^n \lp \chi_{112222} - \rmi^n  \chi_{113221} \rp \rsq~, ~~~ r = 38~. 
    \end{array}\right.
\end{equation}


\textbf{$\tad = 36$:}
\begin{equation}
\begin{aligned}
    G &= \frac{\rmi^m}{32} \lsq (1+\rmi) \chi_{\ll_1} + \rmi^n \lp (\tfrac 12 \pm \rmi) \chi_{\ll_2} + \rmi^n (1-\rmi) \chi_{\ll_3} \rp \rsq \\
    (\ll_1, \ll_2, \ll_3) &= \left\{
    \begin{array}{ll}
    (111133, 112222, 113311), ~~ 
    \end{array}\right.
\end{aligned}
\end{equation}
with rank $50$.


\textbf{$\tad = 40$:}
\begin{equation}
\begin{aligned}
    G &= \frac{\rmi^m}{64} \lsq 4 \chi_{\ll_1} \pm 4\rmi \chi_{\ll_2} + (\pm 1 \pm \rmi) \chi_{\ll_3} \rsq  \\
    (\ll_1, \ll_2, \ll_3) &= \left\{
                \begin{array}{ll}
                  (111133, 111313, 112222), ~~ (111133, 111313, 122122), \\
                  (111133, 111313, 122221), ~~ (111133, 111313, 222121), \\
                  (111133, 111313, 222112), ~~ (111133, 131311, 112222), \\
                  (111133, 331111, 112222), ~~ (111133, 113311, 221212),
                \end{array}~
              \right.
\end{aligned}
\end{equation}
with ranks $44$, $52$, $52$, $62$, $56$, $55$, $60$, $62$.

\begin{equation}
\begin{aligned}
    G &= \frac{\rmi^m}{64} \left\{
    \begin{array}{ll}
         \lsq 2 (1+\rmi) \chi_{\ll_1} + 2 \rmi^n \chi_{\ll_2} + \rmi^p (1+\rmi) \chi_{\ll_3}  \rsq \\
         \lsq 2 (1+\rmi) \chi_{\ll_1} +  \rmi^n (1+\rmi) \chi_{\ll_2} + 2 \rmi^p  \chi_{\ll_3}  \rsq 
    \end{array}\right.  \\
    (\ll_1, \ll_2, \ll_3) &= \left\{
                \begin{array}{ll}
                  (111223, 112222, 121222), ~~ (111223, 122122, 212122), \\
                  (111223, 112222, 122122), ~~ (111223, 112222, 221122), \\
                  (111223, 112222, 222112), ~~ (111223, 122122, 122212), \\
                  (111223, 122122, 212212), ~~ (111223, 122122, 222112), \\
                  (111223, 112222, 122221), ~~ (111223, 112222, 221221), \\
                  (111223, 112222, 222121), ~~ (111223, 122122, 122221), \\
                  (111223, 122122, 212221), ~~ (111223, 122122, 222121), \\
                  (111223, 122122, 222211), ~~ (111223, 122221, 222112), \\
                  (111223, 222112, 222121), ~~ (111223, 122221, 212221), \\
                  (111223, 122221, 222121), ~~ (111223, 222121, 222211),
                \end{array}~
              \right.
\end{aligned}
\end{equation}
with ranks $42$, $48$, $46$, $54$, $64$, $52$, $64$, $56$, $44$, $52$, $58$, $46$, $56$, $54$, $64$, $60$, $56$, $46$, $54$, $62$. 

\begin{equation}
    G = \left\{
    \begin{array}{ll}
         \frac{\rmi^m}{64} \lsq 4 \chi_{111133} + (-1)^s 4\rmi \chi_{113311} \pm (1 + (-1)^s \rmi) \chi_{112222} \rsq~, ~~~ r = 48,  \\[0.2cm]
         \frac{\rmi^m}{64} \lsq 4 \chi_{111133} + (-1)^s 4\rmi \chi_{113311} \pm (1 - (-1)^s \rmi) \chi_{112222} \rsq~, ~~~ r = 50~. 
    \end{array}\right.
\end{equation}

\noindent Many of these solutions are above the upper bound $\bII<3/2$ and are all smaller than 3.

\bibliographystyle{JHEP}
\bibliography{Bibliography}

\end{document}